\documentclass[preprint,12pt]{elsarticle}

\usepackage{graphicx}
\usepackage{amssymb}
\usepackage{amsmath}
\usepackage{amstext} 
\usepackage{xcolor}
\usepackage[version=4]{mhchem}
\usepackage{siunitx}
\usepackage{longtable,tabularx}
\usepackage{textcomp}
\usepackage{subcaption}
\usepackage{tikz}
\usepackage{booktabs}
\usetikzlibrary{positioning} 
\usetikzlibrary{shapes.geometric}
\usetikzlibrary{calc}
\usetikzlibrary{arrows.meta}

\usepackage{lineno}

\journal{Computers \& Fluids}

\newcount\ndots
\def\drawline#1#2{\raise 2.5pt\vbox{\hrule width #1pt height #2pt}}

\def\cqfd{\unskip\kern 6pt\penalty 500\hbox{\vrule\vbox to 4pt{\hrule width 4pt\vfill\hrule}\vrule}}

\def\plus{\raise 1.25pt \hbox{$\scriptstyle +$}\nobreak\ }
\def\x{\raise 1.25pt \hbox{$\scriptstyle \times$}\nobreak\ }

\begin{document}

\begin{frontmatter}


\title{Towards scalable surrogate models based on Neural Fields for large scale aerodynamic simulations}


\author[label1,label2,label3]{Giovanni Catalani}
\author[label1]{Jean Fesquet}
\author[label2]{Xavier Bertrand}
\author[label2]{Frédéric Tost}
\author[label1]{Michael Bauerheim}
\author[label1,label2,label3]{Joseph Morlier}
\address[label1]{ISAE-SUPAERO, Toulouse, France}
\address[label2]{AIRBUS, Toulouse, France}
\address[label3]{ICA, Université de Toulouse, France}

\begin{abstract}
This paper introduces a novel surrogate modeling framework for aerodynamic applications based on Neural Fields. The proposed approach, MARIO (Modulated Aerodynamic Resolution Invariant Operator), addresses non parametric geometric variability through an efficient shape encoding mechanism and exploits the discretization-invariant nature of Neural Fields. It enables training on significantly downsampled meshes, while maintaining consistent accuracy during full-resolution inference. These properties allow for efficient modeling of diverse flow conditions, while reducing computational cost and memory requirements compared to traditional CFD solvers and existing surrogate methods.
The framework is validated on two complementary datasets that reflect industrial constraints. First, the AirfRANS dataset consists in a two-dimensional airfoil benchmark with non-parametric shape variations. Performance evaluation of MARIO on this case demonstrates an order of magnitude improvement in prediction accuracy over existing methods across velocity, pressure, and turbulent viscosity fields, while accurately capturing boundary layer phenomena and aerodynamic coefficients. Second, the NASA Common Research Model features three-dimensional pressure distributions on a full aircraft surface mesh, with parametric control surface deflections. This configuration confirms MARIO's accuracy and scalability. Benchmarking against state-of-the-art methods demonstrates that Neural Field surrogates can provide rapid and accurate aerodynamic predictions under the computational and data limitations characteristic of industrial applications.
\end{abstract}


\end{frontmatter}


{
\noindent

}

\section{Introduction}
\label{sec:intro}
Computational Fluid Dynamics (CFD) serves as a cornerstone of modern simulation in aerospace design, providing high-fidelity characterization of complex flows while reducing dependence on costly experimental testing. In the aeronautical sector, high-fidelity numerical simulations of aircraft aerodynamics are primarily based on the Reynolds-averaged Navier-Stokes (RANS) \cite{pope2001turbulent} equations, allowing for the characterization of the aerodynamic loads, stability characteristics, and overall performance across different configurations. 
During early design phases, CFD is an invaluable tool for evaluating novel geometries and validating innovative configurations.
However, the computational expense associated with high-fidelity CFD can become a significant bottleneck in preliminary design, where numerous iterations are essential for identifying optimal configurations and understanding correlations between complex flow phenomena. This computational burden motivates the development of surrogate models and low-fidelity methods capable of approximating complex flows at reduced cost, while maintaining acceptable accuracy for design space exploration. \\
Proper Orthogonal Decomposition (POD) \cite{berkooz1993proper} has emerged as a popular approach for its interpretable framework, enabling linear dimensionality reduction of high-dimensional non linear systems.
In practice, POD based surrogates are frequently paired with interpolation techniques such as radial basis functions or Gaussian Processes \cite{saves2024smt}, and have seen widespread use in both reduced-order modeling and data-driven applications \cite{fossati2015evaluation,ghoreyshi2024evaluation}. The main limitations of POD relate to the impossibility to handle cases with non uniform grids and geometric variations, as well as the poor performance in capturing strong non-linearities such as shock waves due to its linear formulation \cite{catalani2023comparative}. Recent work has dealt with the extension of POD to circumvent these limitations, such as using mesh morphing for geometry adaptation \cite{casenave2024mmgp}, snapshot clustering \cite{dupuis2018aerodynamic},  or POD basis enrichment with discontinuous functions \cite{ghoreyshi2024evaluation}. 

The emergence of Deep Learning (DL) and Artificial Neural Networks (ANNs) has multiplied data-driven approaches to fluid dynamics modeling and simulation \cite{vinuesa2024opportunities}. Their capacity to extract meaningful representations from high-dimensional systems has enabled significant advances across multiple applications: reduced-order modeling, flow control \cite{martin2024vortex}, hybrid approaches that combine both high-fidelity and data-driven methods for turbulence closure modeling \cite{xu2022pde,de2024space}, and solver acceleration \cite{ajuria2020towards,agarwal2025accelerating}. For applied aerodynamic tasks, ANNs can be leveraged for integral loads estimation \cite{DIASRIBEIRO2023105949}, pressure distribution calibration \cite{bertrand2019wing}, and within multi-disciplinary optimization frameworks \cite{wei2024deepgeo}.
These tasks require the development of reliable surrogates of the underlying flowfield, which, if accurately modeled, can eliminate the computational burden of repeated high-fidelity simulations.
Convolutional Neural Networks (CNNs) represent one of the earliest successful deep learning architectures, owing to their effective inductive bias toward local receptive fields and hierarchical feature extraction \cite{lecun2015deep}. CNN-based architectures, particularly multiscale versions (such as UNet \cite{ronneberger2015u}), have demonstrated promising results in predicting scalar and vector flow fields for both 2D and 3D fluid dynamics test cases \cite{duru2022deep,catalani2023comparative,fesquet2024application}. However, these approaches face inherent limitations in aerodynamic applications. CNNs require structured grid-like input data, necessitating interpolation from the unstructured meshes commonly employed in CFD practice \cite{catalani2023comparative}. This interpolation step inevitably compromises accuracy, particularly in regions with steep gradients. Additionally, the increasing resolution requirements of 3D aerodynamic simulations introduce substantial computational demands that constrain the practical scalability of CNN-based approaches for industrial applications.

By contrast, geometric deep learning, particularly through Graph Neural Networks (GNNs), extends convolutional operations to unstructured grids, where neighborhood definitions are less straightforward \cite{bronstein2017geometric}. GNNs employ message-passing mechanisms to adapt the CNN template of local kernels to graph structures. This approach enables GNNs to handle the unstructured domains typical in fluid dynamics more naturally than image-based methods, as demonstrated by Mesh Graph Networks \cite{pfaff2020learning} in time-dependent fluid problems with moderate mesh sizes. However, defining effective pooling operations on graphs presents significant challenges, limiting information exchange across distant regions of the computational domain. More fundamentally, the network representation of physical phenomena becomes dependent on the specific discretization and graph topology, compromising generalization across different mesh resolutions or configurations \cite{li2024geometry}. For large-scale meshes, information propagation becomes inefficient, and additional problems related to mesh dependence emerge. While multiscale approaches \cite{fortunato2022multiscale, gao2019graph} attempt to mitigate these limitations, they do so at the cost of increased memory requirements and computational complexity, with pooling operations typically remaining mesh-dependent. Despite notable efforts such as dynamic mesh subsampling for aircraft aerodynamics \cite{hines2023graph}, the scalability and discretization invariance of GNN-based surrogates remain significant challenges, particularly for industrial applications involving large meshes.

Operator learning has emerged as a promising paradigm for surrogate modeling by learning mappings between function spaces rather than discretized data points \cite{azizzadenesheli2024neural}. Unlike CNNs and GNNs, Neural Operators learn and treat physical variables as continuous fields, enabling evaluation at any arbitrary point in the domain regardless of the underlying mesh structure. This removes many constraints on discretization. The Fourier Neural Operator (FNO) \cite{li2020fourier} exemplifies this approach by approximating the integral kernel of the underlying operator in a learned latent space using spectral methods. However, FNOs and similar approaches often rely on uniform grids for efficient Fast Fourier Transform operations, limiting their applicability to unstructured meshes typical in aerodynamic simulations. Extensions that handle non-uniform discretizations and geometric variability, such as GINO \cite{li2024geometry}, typically incorporate graph-based components to extend Fourier layers. 
Neural Fields, also known as Implicit Neural Representations (INRs), offer an alternative approach that directly maps spatial coordinates to function values \cite{park2019deepsdf,chen2019learning}, enabling continuous representations of physical fields through neural network parameterizations. Recent works such as CORAL \cite{serrano2024operator} and Aero-nef \cite{catalani2024neural} have demonstrated the effectiveness of this approach for aerodynamic applications, including 3D wing configurations with variable geometries. These methods typically employ an encode-process-decode framework. First, both geometric variations (represented by signed distance function fields) and output physical fields are encoded into compact latent representations. Second, a mapping between the input and output latent encodings is approximated by a separate neural network: the processor model. Despite proving good performance, such approaches face overfitting challenges on small datasets. The separate latent processing stage introduces an optimization that is decoupled from the final output field reconstruction: small regression errors in the latent space can induce larger errors in the decoded reconstructions. 

In this paper, we present MARIO (Modulated Aerodynamic Resolution Invariant Operator), a neural field-based methodology for surrogate modeling of aerodynamic flows that effectively addresses the challenges of geometric variability and scalability to large meshes. 
Compared to previous Neural Fields architectures based on the encode-process-decode framework, MARIO employs an encode-decode pipeline, described in Section \ref{sec:Model Archtecture}, where the encoded geometric representation is directly used by a downstream decoder to predict physical fields. This architectural simplification yields substantial improvements in accuracy for data-scarce scenarios, while reducing model complexity and mitigating overfitting. The direct coupling between geometric representation and physical field prediction proves particularly effective when training data is limited, a common constraint in industrial aerodynamic applications where high-fidelity simulations are computationally expensive.
The encoder-decoder framework is further simplified into a decoder-only architecture when the geometric variability is fully parameterized (e.g., control surface deflections). In this case, the geometric descriptors are directly processed by the decoder network.
Unlike GNNs that rely on local message passing or standard Neural Operators that struggle with complex geometries, MARIO effectively encodes shape information in a compact latent representation that directly influences physical field predictions.
Furthermore, output physical fields are represented in a discretization-invariant manner, mapping spatial coordinates directly to flow variables independently of any specific mesh structure. This property enables a significant computational advantage: training can proceed on significantly downsampled domains, while inference can be performed at full resolution without degradation in predictive accuracy. This contrasts with mesh-dependent approaches where prediction quality typically varies with sampling resolution. \\
In Section \ref{sec:results}, we validate our framework on two complementary datasets that reflect realistic industrial constraints: the AirfRANS dataset \cite{Bonnet_2023} featuring 2D airfoils with non-parametric geometric variations in incompressible flow conditions, providing both surface and volumetric predictions; and the NASA Common Research Model, a 3D full-aircraft configuration with parametric control surface deflections in transonic flow regimes, focusing on surface pressure predictions. Both utilize industrial-scale meshes with limited training data. MARIO consistently outperforms state-of-the-art GNNs and operator-learning baselines while requiring reduced computational resources, establishing its effectiveness for surrogate modeling under tight data constraints.

\section{Methodology}
\label{sec:methodology}
%

\subsection{Problem Statement}
We consider the development of surrogate models for steady-state fluid dynamics simulations on meshed domains involving geometric variability. Let $\Omega \subset \mathbb{R}^d$ denote the physical domain, with $d$ representing the spatial dimension (typically $d=2$ or $d=3$).
The governing physics can be described by a system of steady-state partial differential equations with appropriate boundary conditions. Let $u(x):\Omega \to  \mathbb{R}^{d_u}$ represent the solution field (e.g., pressure, velocity components, or temperature and $d_u$ its dimension) of the boundary value problem, defined as:

$$\begin{cases}
\mathcal{L}(x, u(x); \mu) = 0, & x \in \Omega \\
\mathcal{B}(x, u(x); \mu) = 0, & x \in \partial\Omega
\end{cases}$$
where $\mathcal{L}$ and $\mathcal{B}$ are respectively the PDE operator (such as the Reynolds-averaged Navier-Stokes equations) and the boundary conditions operator, while $\mu \in \mathbb{R}^{d_\mu}$ denotes global parameters including operating conditions (e.g., angle of attack, Mach number, Reynolds number). 
In general, we are interested in approximating the mapping $\mathcal{G}: (\mu, \Omega) \mapsto u$, with a parameterized map $\mathcal{G}_\theta: (\mu, \Omega) \mapsto u$, where $\theta \in \Theta$ represents a set of learnable parameters. 
In practice, the domain $\Omega$ is discretized as a mesh $\mathcal{M}$, and we wish to learn $\mathcal{G}_\theta$ from a dataset $\mathcal{D} = \{(\mu_i, \mathcal{M}_i), u_i\}_{i=1}^M$, where each tuple consists of a mesh $\mathcal{M}_i$ with nodes $\mathcal{V}_i$ representing the discretized domain $\Omega_i$, parameters $\mu_i$, and the corresponding solution field $u_i$ computed by a numerical solver. The optimal parameters $\theta^\dagger$ are estimated by solving the optimization problem:
\begin{equation}
\theta^\dagger = \arg\min_{\theta \in \Theta} \frac{1}{M} \sum_{i=1}^M \sum_{x \in \mathcal{V}_i} \|u_i(x) - [\mathcal{G}_\theta(\mu_i,\mathcal{M}_i)](x)\|^2.
\end{equation}
In this work, we address both parametric and non-parametric geometric variability. In the first case, the domain geometry is explicitly described by a known set of parameters $\mu_{\text{geom}}$ (e.g., control surface deflections or airfoil shape parameters), while in the latter, an explicit parametrization is not known a priori. Therefore, a learnt geometric parameterization is obtained by fitting the Signed Distance Function (SDF) $sdf(x)$, defined as the minimum distance from point $x$ to the boundary $\partial\Omega$, with negative values inside the domain and positive values outside:
\begin{equation}
sdf(x) = s(x) \cdot \min_{y \in \partial\Omega} \|x - y\|, \quad s(x) = \begin{cases} -1, & \text{if } x \in \Omega \\ 1, & \text{if } x \not\in \Omega. \end{cases}
\end{equation}
Note that in aerodynamic surrogate modeling applications, the main interest can often be limited to predicting physical quantities only on the domain boundary $\partial\Omega$ rather than the entire volume. For example, the pressure and skin friction distributions on the aircraft surfaces are sufficient to directly determine the aerodynamic forces and moments. In these cases, the observed solution field $u_i$ may be restricted to the boundary $\partial\Omega_i$. This remark is important to understand the experiments presented in the following sections, but does not drastically change the proposed methodology.

\subsection{MARIO framework}
\label{sec:Model Archtecture}
To address the surrogate modeling problem formulated in the previous section, we develop an approach based on Neural Fields that effectively handles both parametric and non-parametric geometric variability. Our framework aims to learn the mapping from operating conditions and geometric configurations to flow fields, scaling to large meshes in 2D and 3D. In the following sections, we first introduce the key concepts of Conditional Neural Fields, then describe how they are applied to the specific cases of parametric and non-parametric geometric variability. We then detail the geometry encoding process and the implementation specifics of our neural network architecture, which together form the complete MARIO surrogate modeling framework.

\subsubsection{Conditional Neural Fields}
\begin{figure}
    \centering
    \includegraphics[width=0.9\linewidth]{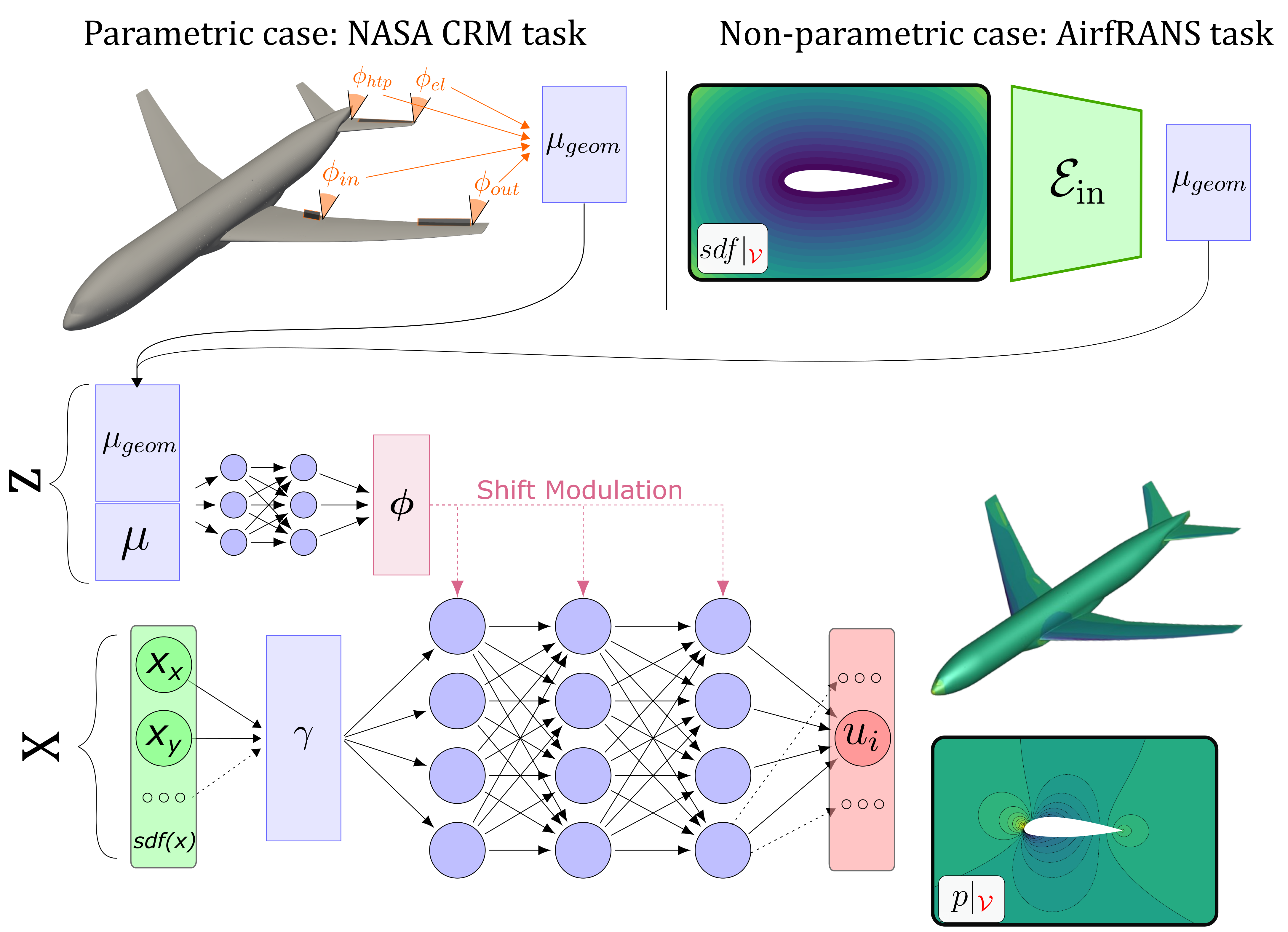}
    \caption{MARIO's conditional neural field architecture. The framework handles two types of geometric variability: (1) Parametric case (e.g., NASA CRM), where $\mu_{\text{geom}}$ is directly obtained from explicit parameters like control surface deflections; and (2) Non-parametric case (e.g., AirfRANS), where $\mu_{\text{geom}}$ is learned by encoding SDF fields as described in Section \ref{sec:Geometry Encoding}. The conditions $z$ are formed by concatenating operating conditions $\mu$ (e.g., $M$, $\alpha$, $Re$) with $\mu_{\text{geom}}$. Spatial coordinates $x_x,x_y,..$ and optional auxiliary fields (e.g., normals, SDF) are processed through Fourier feature encoding ($\gamma$). The hypernetwork $h_\psi$ generates modulation vectors $\boldsymbol{\phi}$ from $z$, which are applied to each layer of the main network to condition the output.}
    \label{fig:mario_overview}
\end{figure}

Neural Fields, or Implicit Neural Representations, are parametric functions $f_\theta: \mathbb{R}^d \rightarrow \mathbb{R}^{d_u}$ parameterized by weights $\theta$, mapping inputs $x \in \mathbb{R}^d$ (coordinates and optionally additional fields) to target features $u \in \mathbb{R}^{d_u}$. By design, Neural Fields model data as continuous functions, offering several advantages: as $f_{\theta}(x)$ can be queried at any $x$, they can represent objects (images, shapes, physical fields..) at levels of discretization that are independent of the training resolution. Additionally, they offer memory-efficient representations of data through the parameters $\theta$, whose dimension scales with the size of the network rather than the number of discrete samples (for instance the number of pixels in an image) as in the case of array-based representations. These properties can explain the scalability of Neural Fields to handle complex, high-dimensional tasks. For example, a Neural Field parametrization of a single image (or a shape) can be obtained by fitting the parameters to map the pixel coordinates to the pixel values. While this has interesting applications in scene or shape representations, for surrogate modeling, an entire family of solutions should be represented using a shared parametrization. Conditional Neural Fields extend this framework by introducing a sample-specific conditioning variable $z \in \mathbb{R}^{d_z}$ (together with a shared parametrization $\theta$) that carries the information of a particular data sample. This conditioning mechanism allows a single neural network architecture to represent multiple distinct signals by varying $z$, making it well suited for the surrogate modeling task in which different geometries and flow conditions must be modeled.
In the simplest form, conditioning can be implemented by concatenating the conditioning variable with the input features \cite{park2019deepsdf}. However, more efficient conditioning mechanisms exist, where global parameters (conditions) are processed differently compared to local parameters (coordinates). MARIO is implemented as a modulated conditional Neural Field:
\begin{align}
f_{\theta,\phi}(x) &= W_L(\eta_{L-1} \circ \eta_{L-2} \circ \cdots \circ \eta_1 \circ \gamma(x)) + b_L \\
\eta_l(\cdot) &= \text{ReLU}(W_l(\cdot) + b_l + \phi_l(z))
\end{align}
where layer-wise modulation vectors are obtained as a function of the conditioning variable through a fully connected hypernetwork $h$ (parametrized by $\psi$): $\phi_l(z) = [h_{\psi}(z)]_l \in \mathbb{R}^{d_l}$. The modulations are layer-specific and operate by shifting the outputs of the intermediate layer, before the application of the non-linear activation function (ReLU in this case). It follows that modulations and activations $\eta_l$ share the same dimension $d_l$. The main network parameters $\theta$ are shared for all samples and consist of the weights and biases matrices $W_l,b_l$. 
To address the spectral bias inherent in neural networks, we employ Fourier feature encoding for the input coordinates:
\begin{equation}
\gamma(x) = [\cos(2\pi \mathbf{B}x), \sin(2\pi \mathbf{B}x)]
\end{equation}
where $\mathbf{B} \in \mathbb{R}^{m \times d}$ contains frequency vectors sampled from a Gaussian distribution $\mathcal{N}(0,\sigma)$. In our implementation, we use a single scale $\sigma$ for simplicity, although multi-scale approaches with different $\sigma$ values could provide finer control over the kernel bandwidth for different output signals \cite{catalani2024neural}. Neural Tangent Kernel (NTK) \cite{jacot2018neural} theory demonstrates that standard MLPs learn frequencies at rates decreasing exponentially with frequency, but Fourier feature encoding \cite{tancik2020fourier} creates a stationary NTK that enables better convergence to high-frequency components of the target function.
The sample diversity information is carried by the conditioning variable $z$: in MARIO, this is obtained by concatenating the operating conditions $\mu$ and the geometric parametrization $\mu_{geom}$: $z=[\mu,\mu_{geom}]$.
In Figure \ref{fig:mario_overview}, an overview of MARIO architecture is displayed. 
The geometric parameters $\mu_{geom}$, in particular, can be explicitly known (for instance, as control surface deflections) or may not be directly available.
In the latter case, a compact parameterization of the domain geometry can be obtained by encoding the Signed Distance Function fields defined over the shape of the domain boundary.
For this encoding task, a separate neural field $f_{\theta_{in},\phi_{in}}$ where $\phi_{in}=h_{\psi_{in}}(z_{in})$ is used to find the optimal conditioning variables $z_{in}=\mu_{geom}$ that minimize the reconstruction error of the signed distance function fields:

\begin{equation}
    z_{in} = \arg\min_z \mathcal{L}(f_{\theta_{in},\phi_{in}}(x), sdf), \quad \phi_{in}=h_{\psi_{in}}(z_{in}).
\end{equation}
This optimization problem, known as auto-encoding ($z_{in}=\mathcal{E}_{in}(sdf)$) \cite{park2019deepsdf}, is further detailed in Section \ref{sec:Geometry Encoding}. In the next section, the training procedure is explained, differentiating between the parametric and non-parametric geometric variability cases.

\subsection{Training Methodology}
\label{sec:Training}
For surrogate modeling tasks, we employ conditional neural fields to predict physical fields as a function of domain geometry and operating conditions. The training procedure depends on how geometric information is represented, as illustrated in Figure \ref{fig:mario_overview}:

\begin{enumerate}
    \item \textbf{Parametric Geometric Variability}: When explicit geometric parameters $\mu_{\text{geom}}$ are available (as in the NASA CRM case with control surface deflections), these are directly used to form the conditioning variable $z=[\mu_{\text{geom}}, \mu]$. Training involves optimizing a single set of neural field parameters $\theta$ and hypernetwork parameters $\psi$ by minimizing:
    \begin{equation}
    \mathcal{L} = \sum_{i=1}^{n_{\text{tr}}} \sum_{x \in \mathcal{V}_i} \|u_i(x) - f_{\theta,\phi}(x)\|^2, \quad \phi=h_{\psi}([\mu_{\text{geom}}^i, \mu^i])
    \end{equation}

    \item \textbf{Non-parametric Geometric Variability}: When explicit geometric parameters are unavailable (as in the AirfRANS dataset with arbitrary airfoil shapes), geometric variability is captured through the encoding process described in the previous section. Training proceeds in two stages:
    
    (i) First, for each geometry, we obtain the geometric parametrization $\mu_{\text{geom}}^i = \mathcal{E}_{in}(sdf_i)$ by autoencoding the signed distance function fields,
    while simultaneously optimizing the shared parameters $\theta_{in}$ and $\psi_{in}$ across all training geometries.
    
    (ii) Then, using these learned geometric representations, we optimize the output neural field parameters $\theta$ and $\psi$ by minimizing:
    \begin{equation}
    \mathcal{L} = \sum_{i=1}^{n_{\text{tr}}} \sum_{x \in \mathcal{V}_i} \|u_i(x) - f_{\theta,\phi}(x)\|^2, \quad \phi=h_{\psi}([\mu_{\text{geom}}^i, \mu^i])
    \end{equation}
\end{enumerate}
This approach allows MARIO to handle both parametric and non-parametric geometric variations within a unified framework. Figure \ref{fig:mario_overview} illustrates both cases: the top row shows the direct conditioning approach for parametric geometric variability (NASA CRM), while the bottom row depicts the two-stage process for non-parametric geometric variability (AirfRANS).

\subsubsection{Geometry Encoding}
\label{sec:Geometry Encoding}
In this section, the encoding process $\mathcal{E}_\text{in}({sdf}_{i})$ used to obtain latent representations of geometry in the non-parametric case is described. For each geometry's signed distance function (SDF), a meta-learning optimization procedure based on CAVIA \cite{zintgraf2019fast} adapts a shared neural network to represent different shapes. Given the shared network parameters $\theta_{in}$ and hypernetwork parameters $\psi$, the latent representation $z_{in}^{(k)}$ for geometry $i$ is obtained by solving an optimization problem:
\begin{align}
    z_{in}^{(0)} &= 0 \\
    z_{in}^{(k+1)} &= z_{in}^{(k)} - \alpha \nabla_{z_{in}^{(k)}} \mathcal{L}_{in}(f_{\theta_{in},\phi_{in}}(x), sdf_i), \quad \text{for } 0 \leq k \leq K-1
\end{align}
 where \quad $\phi_{in}=h_{\psi}(z_{in}^{(k)})$, $\alpha$ is the inner loop learning rate, and $K$ is the number of optimization steps (typically set to 3). The loss $\mathcal{L}_{in}$ is a mean square error function, measuring the reconstruction error between the true SDF field and its prediction $\widehat{sdf}$ over a sampling grid $\mathcal{V}$ defined on the input domain.
During training of the geometry encoder, the optimization of geometric latent vectors $z_{in}$ is performed as an internal loop within the outer loop of the global parameters optimization: at each epoch for each training batch, the global weights ($\theta_{in},\psi$), fixed during the $K$ inner loop steps, are then updated using standard gradient descent. The resulting gradients then flow through the inner steps. This results in a second-order optimization algorithm, with increased memory requirements (as the backpropagation graph needs to be stored across the inner loop steps) but improved training stability \cite{serrano2024operator,dupont2022data}. This approach offers several practical advantages: first, it requires only a few ($K$) gradient steps to encode a new geometry at inference time, significantly reducing computational overhead. Second, since latent codes are re-initialized to zero and updated in just $K$ steps, their magnitude remains bounded and controlled by the inner learning rate, providing effective latent space regularization. As a consequence, overfitting is mitigated and generalization to unseen geometries is improved.

\begin{figure}
    \centering
    \includegraphics[width=0.9\linewidth]{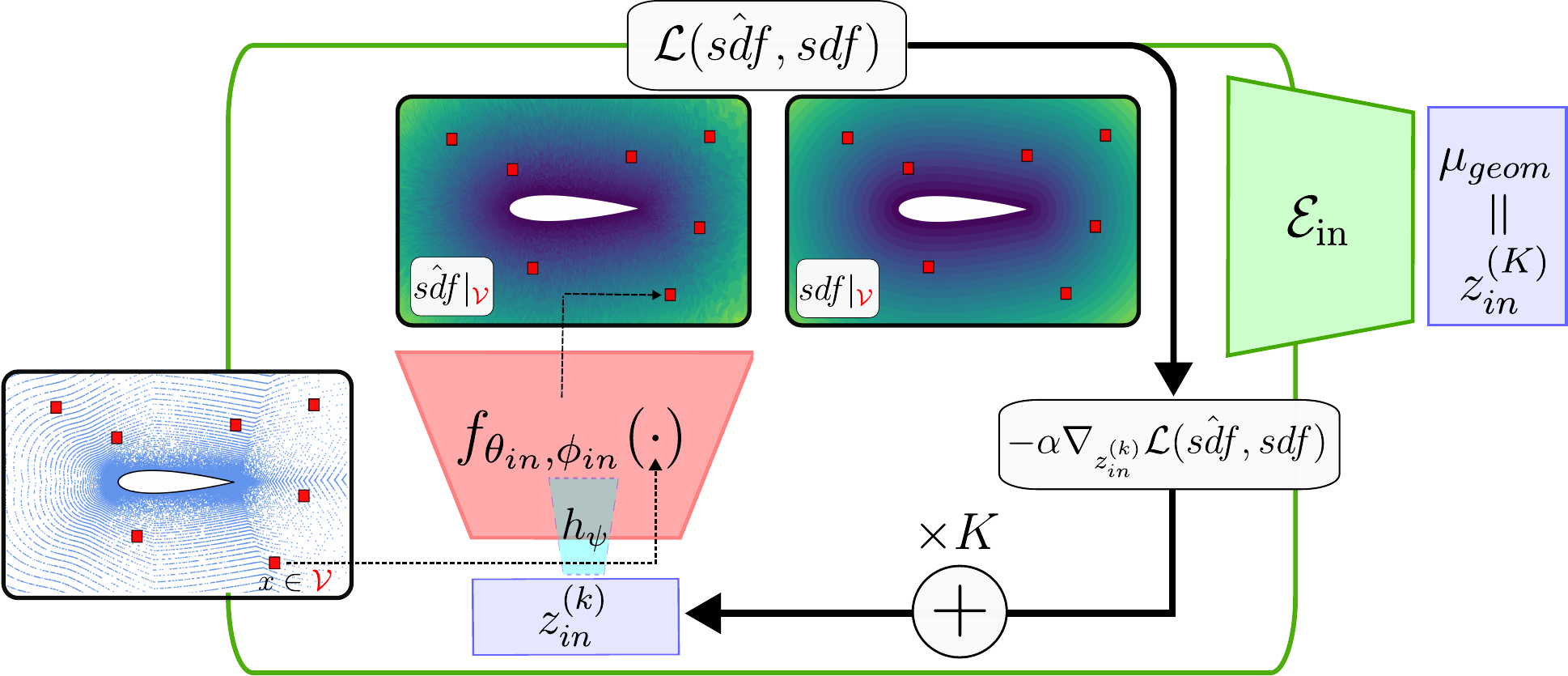}
    \caption{Geometry encoding process used in MARIO for non-parametric geometric variability. The neural field $f_{\theta_{in},\phi_{in}}$ with modulation vectors from hypernetwork $h_{\psi}$ takes points (in red) $x \in \mathcal{V}$ and the current latent code $z_{in}^{(k)}$ to predict the SDF field. The loss between the true and predicted SDF fields is used to compute the gradient with respect to $z_{in}^{(k)}$, which is then used for $K$ gradient descent steps. The final latent code $z_{in}^{(K)}$ becomes the geometric descriptor $\mu_{geom}$.}
    \label{fig:geometry_encoding}
\end{figure}

\section{Datasets and experimental settings}
\label{sec:db}

This section presents the two test cases used to validate the proposed surrogate modeling framework. Both datasets include significant challenging features for surrogate modeling in aerodynamics:

\begin{itemize}
    \item \textbf{Large-scale meshes:} both datasets feature high-resolution CFD meshes of approximately 200,000 nodes per sample for AirfRANS (both volumetric and surface) and 450,000 surface nodes for the NASA CRM.
    \item \textbf{Data scarcity:} we focus on limited training data scenarios that reflect real-world constraints in industrial applications: 200 training samples for AirfRANS and 100 samples for the NASA CRM.
\end{itemize}
These datasets also present distinct modeling challenges. The AirfRANS dataset involves 2D airfoils with non-parametric shape variations, requiring the complete MARIO pipeline (geometry encoder and output decoder) as described in Section~\ref{sec:Geometry Encoding}. In contrast, the NASA CRM dataset features a fixed 3D aircraft geometry with parametric control surface deflections, allowing direct usage of the output neural field without the geometry encoding stage.
Together, on these test cases, the proposed approach can be evaluated on both 2D and 3D configurations, volumetric and surface-only predictions, and parametric versus non-parametric geometric variations. Table~\ref{tab:datasets_comparison} summarizes the key characteristics of both datasets and the corresponding modeling tasks. We provide the full implementation of MARIO and all the experiments on the AirfRANS dataset at \texttt{https://github.com/giovannicatalani/MARIO} to facilitate reproducibility and extension of our work.

\begin{table}[t]
    \centering
    \caption{Comparison of dataset characteristics and modeling tasks}
    \label{tab:datasets_comparison}
    \begin{tabular}{lcc}
        \toprule
        \textbf{Characteristic} & \textbf{AirfRANS} & \textbf{NASA CRM} \\
        \midrule
        Geometry & 2D & 3D  \\
        Geometric variation & Non-parametric & Parametric \\
        N. Mesh nodes & $\sim$200000 nodes & 454000 nodes \\
        Train-Val/Test split & 200/200 & 105/44 \\
        Domain & Volume + surface & Surface\\
        Output variables & $u_x$, $u_y$, $p$, $\nu_t$ & $C_p$ \\
        Neural Field approach & Encoder-Decoder & Decoder only \\
        \bottomrule
    \end{tabular}
\end{table}

\subsection{AirfRANS Dataset}
\subsubsection{Dataset and Task Description}
\begin{figure}[h]
    \centering
    \includegraphics[width=\textwidth]{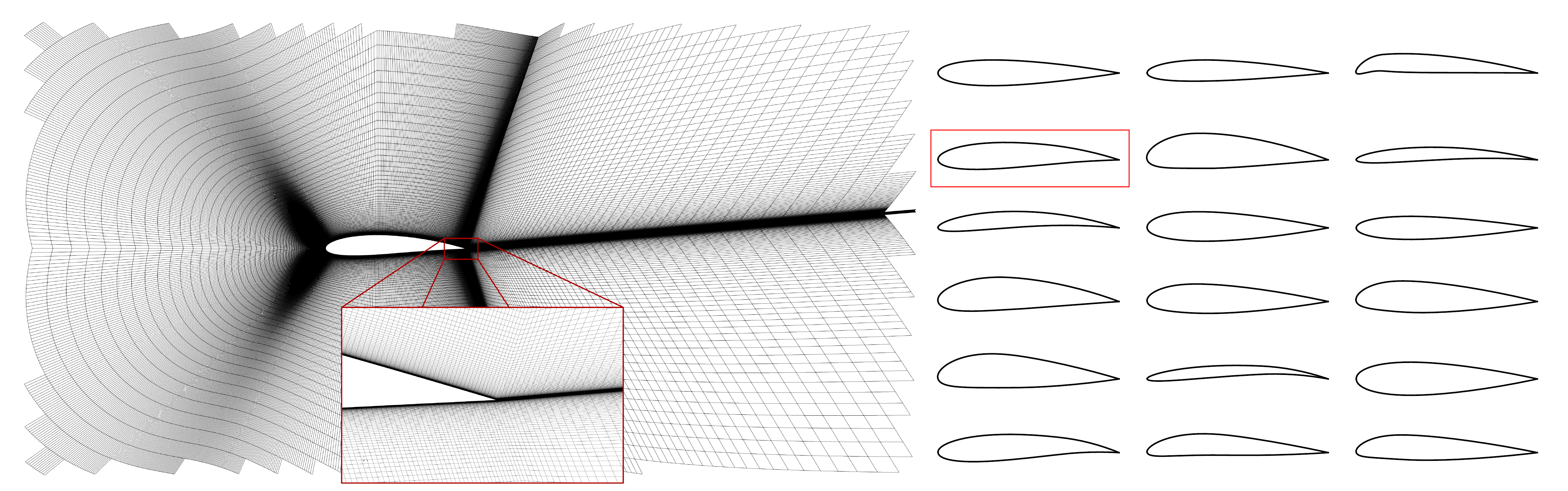}
    \caption{Left: Example of the computational mesh for a single airfoil geometry in the AirfRANS dataset, with a zoomed view highlighting the refined mesh resolution near the airfoil surface. Right: Representative sample of diverse airfoil geometries from the dataset.}
    \label{fig:mesh_AirfRANS}
\end{figure}

The AirfRANS dataset \cite{Bonnet_2023}  presents a comprehensive benchmark for computational fluid dynamics (CFD) surrogate modeling specifically designed to test algorithmic performance in aerodynamic applications. It consists of steady-state RANS simulations of subsonic, incompressible flow around two-dimensional airfoil geometries spanning a wide range of shapes and operating conditions. The dataset reflects industrial challenges through its use of highly refined meshes and relative data scarcity.
Additionally, the controlled training and test dataset split settings allow for fair model comparisons.
The numerical solutions of the RANS equations have been obtained using OpenFOAM's \cite{jasak2009openfoam} incompressible solver with the Spalart-Allmaras (SA) turbulence model \cite{spalart1992one}, which introduces a single transport equation for a modified eddy viscosity $\tilde{\nu}$ from which the turbulent viscosity $\nu_t$ is then computed. Each case employs a structured C-grid hexahedral mesh with significant refinement near the airfoil surface to accurately resolve boundary layer phenomena.
The different airfoil geometries systematically derived from the NACA 4-digit and 5-digit series parameterizations.
A subset of sample geometries, as well as a visualization of the mesh used for a specific geometry, is provided in Figure \ref{fig:mesh_AirfRANS}.
Flow conditions span Reynolds numbers from 2 to 6 million and angles of attack from -5° to 15°, representing typical subsonic flight regimes.
For machine learning applications, the dataset is provided in PyTorch Geometric format \cite{gardner2018gpytorch}. Each sample contains input data consisting of node coordinates $x,y$, surface normals $n_x$, $n_y$, signed distance function field $sdf(x,y)$
and freestream velocity components $V_x,V_y$. The output fields include velocity components $u_x,u_y$, pressure $p$, and turbulent viscosity $\nu_t$ fields.
For the Machine Learning task, we focus here on the '\textit{scarce}' settings: 200 simulations are used for training and validation, and 200 simulations are used for testing. This specific train-test split is chosen to reproduce a typical industrial scenario where high-fidelity data is scarcely available, and to ensure model generalization is realistically demonstrated.

\subsubsection{Experimental Settings and Baselines}
In this section, we describe the experimental setup and specific implementation choices for the proposed model on the AirfRANS dataset. As detailed in Section~\ref{sec:Model Archtecture}, for such test cases, involving non-parametric geometric variability, the geometry is separately encoded via its Signed Distance Function (SDF) using an input encoder. 
The proposed model is compared against various state-of-the-art approaches based on Graph Neural Networks and Transformer-based architectures, briefly described in the following paragraphs.
\paragraph{MARIO}
(i) The input encoder is implemented as a Neural Field that maps 2D spatial coordinates to the corresponding SDF of each airfoil geometry. This encoder is built upon a five-layer fully connected network with a hidden dimension of 256, augmented by a single-layer hypernetwork. Fourier Feature encoding is applied on the input features, as 64 Fourier Features are sampled from a Gaussian Distribution of standard deviation $\sigma=1$. The encoding process, described in Section \ref{sec:Geometry Encoding}, produces latent vectors of dimension $d_z=8$, for each sample geometry. 
(ii) The output neural field, acting as a decoder, receives as input a concatenation of the encoded geometry latent vectors and the operating conditions $V_x,V_y$. This is processed by a 3-layer hyper-network and results in layer-wise modulation vectors, that are applied on the activations of the main network (Section \ref{sec:Model Archtecture}). The main network inputs are the spatial coordinates, the signed distance function, the surface normals, and an additional feature-engineered input $\sigma_{bl}$ referred to as boundary layer mask. The main philosophy behind this mask is to facilitate network learning in the near-wall region of the domain, where the boundary layer develops. This mask is derived from the Signed Distance Function (SDF) through a series of transformations that effectively highlight the near-wall region. Starting with the normalized inverse distance from the surface, we apply a thresholding operation with a specified thickness parameter $\tau=0.02$, followed by a squared function to create a parabolic decay profile:
\begin{equation}
\sigma_{bl} = \left(\max\left(0, \frac{\hat{d} - (1-\tau)}{\tau}\right)\right)^2   
\end{equation}
where $\hat{d}$ represents the normalized inverse distance. This formulation creates a smooth field that equals 1 at the airfoil surface and decays quadratically to 0 at a distance $\tau$ from the surface. The mask remains uniform along the chord to minimize computational complexity. This additional input proves particularly effective when simultaneously predicting physical fields with significantly different gradient characteristics, such as velocity and pressure. The specific impact of this boundary layer mask on model accuracy is analyzed in detail in the Results section.
For the presented experiments, the model is trained for 1000 epochs using Adam optimizer \cite{Kingma2014Adam:Optimization} with learning rate $10^{-3}$, on a single NVIDIA A100 GPU, following the procedure in Section~\ref{sec:Training}. In particular, at each training epoch, random subsampling is applied on the discretized fields, and a subset of 16000 points is selected ($\sim 10\%$ of the full mesh). This allows for speed-up training while keeping representative training sample sizes. At inference time, the output neural fields are used to predict the output fields over the full meshes: the discretization invariant properties of the method result in consistent errors over all testing resolutions. 
\paragraph{Baselines}
For baseline comparisons of the performance of the proposed model, we consider a simple Multi-layer-Perceptron (MLP) approach, several Graph Neural Network models: Graph UNET \cite{gao2019graph}, PointNet \cite{qi2017pointnet}, Graph SAGE \cite{hamilton2017inductive}. Additionally, we consider a recently proposed Neural Operator model, named Transolver \cite{wu2024transolver}, which has demonstrated good performance on several large-scale aerodynamic benchmarks. Traditional surrogate modeling approaches such as Proper Orthogonal Decomposition (POD) and Gaussian Process Regression (GPR) were not included in our comparison as they typically face challenges with non-parametric geometric variations. The Mesh Morphing Gaussian Process (MMGP) approach \cite{casenave2024mmgp}, which extends these methods through specialized mesh deformation techniques, has shown strong results for incompressible flows with common mesh topologies, and readers are encouraged to refer to the corresponding publication for more details. However, reproducing its mesh morphing component is not trivial and we decide to focus the baseline comparison with more similar deep learning based methodologies.
\paragraph{Evaluation Metrics}
To assess model performance, we employ metrics that capture both the point-wise accuracy of the predicted flow fields and the global trends in aerodynamic coefficients, as proposed in the original AirfRANS paper \cite{Bonnet_2023}. For the field variables (velocity components, pressure, turbulent viscosity, and surface pressure), we compute Mean Squared Error (MSE) for the normalized output variables averaged over the test set. For the aerodynamic coefficients (specifically the drag coefficient \(C_D\) and lift coefficient \(C_L\)), we report the mean relative errors. In addition, we compute the Spearman's rank correlation coefficients \(\rho_D\) and \(\rho_L\) for the integral aerodynamic coefficients. This metric measures whether both ground truth and predicted coefficients increase or decrease together, and it is more robust to outliers compared to other correlation coefficients as Pearson's. This can be more important when evaluating optimal designs, for example in terms of lift over drag aerodynamic efficiency.


\subsection{NASA Common Research Model}
\subsubsection{Dataset and Task Description}
The NASA CRM is a 3D CFD dataset of RANS simulations over the NASA Common Research Model geometry \cite{rivers2014experimental} (sketched in Figure \ref{fig:mario_overview}). The CFD computations have been performed at DLR \cite{doi:10.2514/1.J061234} and comprise 149 different simulations of the same base geometry with variations in control surface deflections (inboard and outboard ailerons, elevator, and horizontal tailplane) along with changes in flight conditions (Mach number and angle of attack). 
The geometric parameters $\mu_{geom}$ include four control surface deflection angles: inboard aileron deflection ($\phi_{inAil}$) ranging from $-20^{\circ}$ to $20^{\circ}$, outboard aileron deflection ($\phi_{outAil}$) ranging from $-20^{\circ}$ to $10^{\circ}$, elevator deflection ($\phi_{el}$) ranging from $-10^{\circ}$ to $10^{\circ}$, horizontal tailplane incidence ($\phi_{htp}$) ranging from $-2^{\circ}$ to $2^{\circ}$.
These are combined with inflow conditions ($\mu$): Mach number ($M$) ranging from $0.5$ to $0.9$,
Angle of attack ($\alpha$) ranging from $-2^{\circ}$ to $5^{\circ}$.
The computational mesh consists of approximately 43 million points, though only the surface mesh of 454,404 nodes is considered for surrogate modeling of pressure distributions. More details about the dataset generation and the CFD methodology can be found in the original publication \cite{doi:10.2514/1.J061234}.
The goal here is to demonstrate the scalability and flexibility of our model when applied directly to this large 3D case without modifications. As our approach does not require mesh connectivity information or ad-hoc coarsening, the model can be trained at a significantly lower resolution than the full surface mesh and then inferred on the complete surface mesh at test time. The geometric parameters (in this case, the control surface deflection angles) can be directly concatenated with operating conditions and fed into the output neural field to produce predictions of the surface pressure distribution, showcasing the practical advantages of our approach for industrial-scale applications.

\subsubsection{Experimental Settings}
For the NASA CRM dataset, we utilize the parametric geometric variability approach outlined in Section~\ref{sec:Model Archtecture}, employing only the output neural field decoder, as the geometric variations are explicitly parameterized through control surface deflection angles. The model architecture consists of a single neural field that directly maps spatial coordinates to pressure coefficients, conditioned on both geometric parameters ($\mu_{geom}$) and flight conditions ($\mu$).
The conditioning mechanism is implemented through a 3-layer hypernetwork that processes the concatenated input parameters and produces modulation vectors for each layer of the main network, following the same approach described in Section~\ref{sec:Model Archtecture}. The spatial coordinates, along with surface normals, are encoded using Fourier feature encoding with 64 frequency components sampled from a Gaussian distribution with $\sigma=1$. 
At training time, we employ dynamic subsampling where 10,000 points (approximately 2\% of the full surface mesh) are randomly selected at each epoch. This approach significantly reduces computational requirements while preserving the model's ability to learn the underlying pressure distribution patterns. At inference time, predictions can be made on the entire surface mesh without resolution constraints, leveraging the discretization-invariant properties of the neural field formulation.
The model is trained for 10000 epochs using the Adam optimizer with learning rate $10^{-3}$. 

\subsubsection{Evaluation Metrics}
To assess model performance on the NASA CRM dataset, we compute the Mean Squared Error (MSE) of the predicted pressure coefficient ($C_p$) on the test set consisting of 40 unseen configurations. This metric directly quantifies the ability of the model to accurately capture pressure distributions across the aircraft surface, which is the primary quantity of interest for aerodynamic analysis and design. Additionally, we present qualitative visualizations of pressure distributions over the aircraft surface to provide insights into the model's ability to capture complex aerodynamic features such as shock waves.

\section{Results}
\label{sec:results}
%

\subsection{AirfRANS Dataset}
\label{subsec:result_af}

The comparative performance of the MARIO framework against the considered baselines on the AirfRANS dataset is presented in Tables~\ref{tab:model_comparison} and~\ref{tab:aero_coefficients}. As shown in Table~\ref{tab:model_comparison}, MARIO consistently outperforms all baseline models across all field prediction metrics, achieving approximately an order of magnitude improvement in all output fields. Notably, even the strongest baseline models (GraphSAGE for velocity and GUNet for pressure) exhibit errors 8-10 times higher than the proposed approach. 

\begin{table}[htbp]
\centering
\begin{tabular}{lccccc}
\toprule
\textbf{Model} & $\boldsymbol{\bar{u}_x}$ & $\boldsymbol{\bar{u}_y}$ & $\boldsymbol{\bar{p}}$ & $\boldsymbol{\nu_t}$ & $\boldsymbol{\bar{p}_s}$ \\
\midrule
MARIO & \textbf{0.152} & \textbf{0.113} & \textbf{0.24} & \textbf{0.096} & \textbf{0.27} \\
MLP* & 1.647  & \underline{1.451}  & 3.904 & 0.501  & 2.192 \\
PointNet* & 3.111  & 2.776  & \underline{3.294}  & 0.558  & 1.827  \\
GraphSAGE* & \underline{1.457} & 1.454 & 4.696  & 0.611 & 1.945  \\
GUNet* & 1.749  & 1.825 & 3.388 & \underline{0.433}  & \underline{1.473}  \\
Transolver & 2.105 & 2.108 & 4.434 & 0.758 & 2.367 \\
\bottomrule
\end{tabular}
\caption{\textbf{Comparison of different models on the AirfRANS dataset, scarce task}. Values shown are absolute MSE errors on the test set (standardized outputs). Error values are scaled as follows: $\bar{u}_x$, $\bar{u}_y$, and $\bar{p}$ are multiplied by $10^{-2}$; $\nu_t$ and $\bar{p}_s$ are multiplied by $10^{-1}$. The \textbf{bold} values indicate the best (lowest) results, while \underline{underlined} values indicate the second-best results. Models marked with * have values taken from \cite{Bonnet_2023} under identical experimental settings.}
\label{tab:model_comparison}
\end{table}

This performance gap extends to the prediction of integral aerodynamic coefficients, as demonstrated in Table~\ref{tab:aero_coefficients}, where MARIO achieves substantially lower errors in both drag and lift coefficient predictions while maintaining superior Spearman correlation metrics.
The improved performance can be attributed to three key advantages of the neural field approach. 

\begin{figure}[h]
    \centering
    \includegraphics[width=\textwidth]{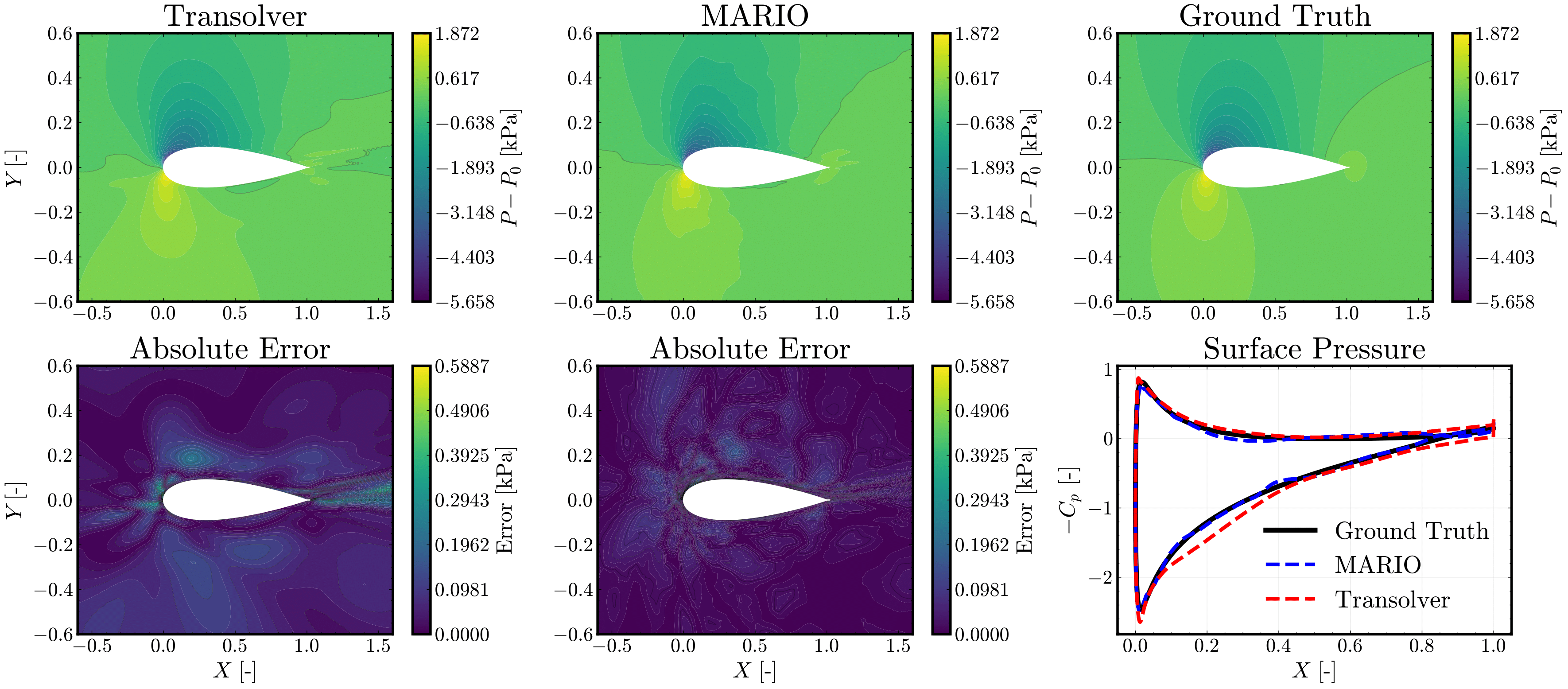}
    \caption{Comparison of pressure field predictions between MARIO and Transolver models for a test sample. Top: Pressure contour plots showing the full field predictions. Bottom: Error contour plots and pressure coefficient ($C_p$) distribution along the airfoil surface. $\alpha=9.39 ^\circ$, $M=0.17$}
    \label{fig:pressure_contour_82}
\end{figure}

\begin{figure}[h]
    \centering
    \includegraphics[width=\textwidth]{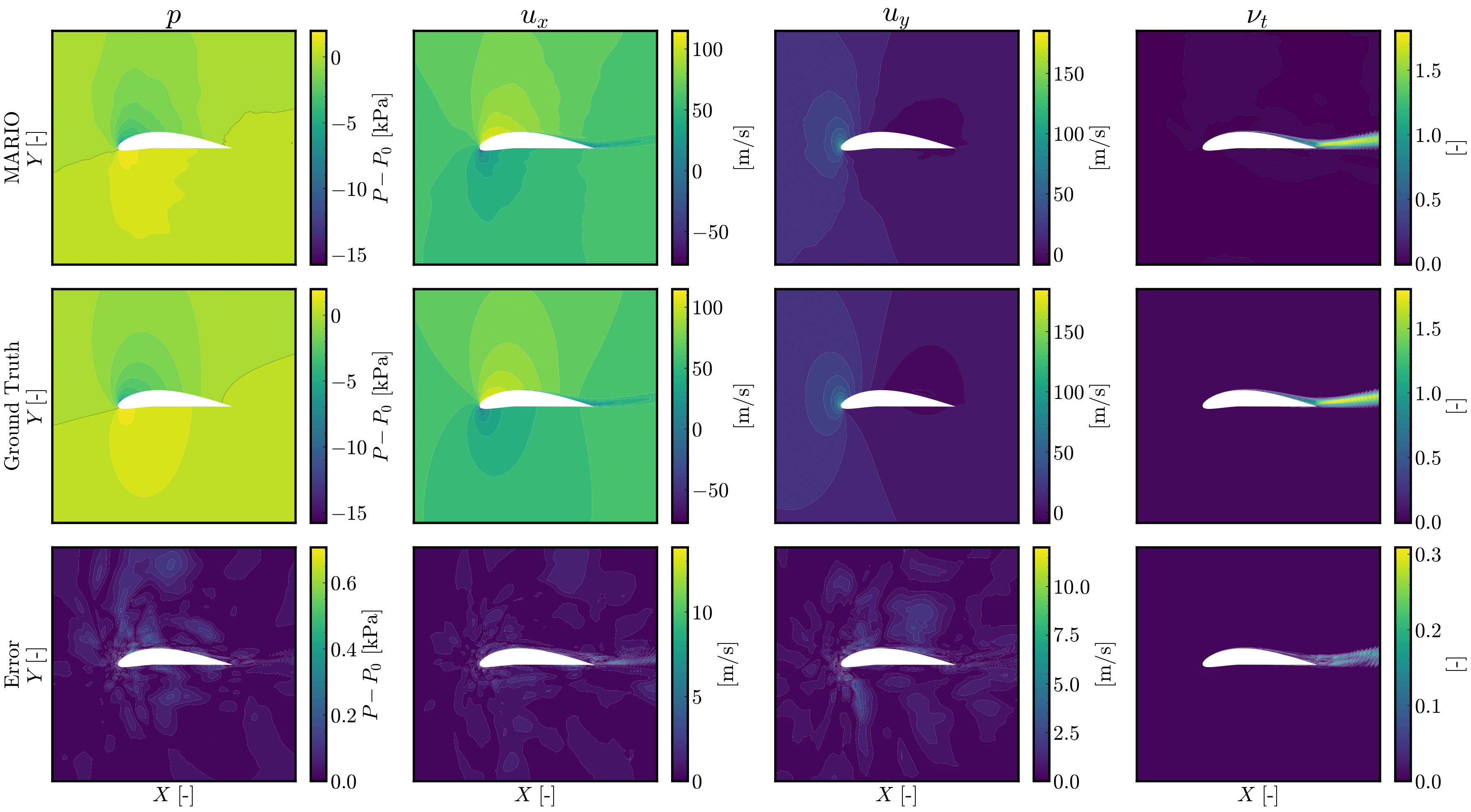}
    \caption{Comparison of all four predicted fields ($u_x$, $u_y$, $p$, $\nu_t$) for an in-distribution airfoil test case. First row: MARIO predictions; Second row: Ground truth CFD results; Third row: Absolute error contour plots. Operating conditions: $\alpha=9.94^\circ$, $M=0.18$.}
    \label{fig:fields_contour_99}
\end{figure}

\begin{figure}[h]
    \centering
    \includegraphics[width=\textwidth]{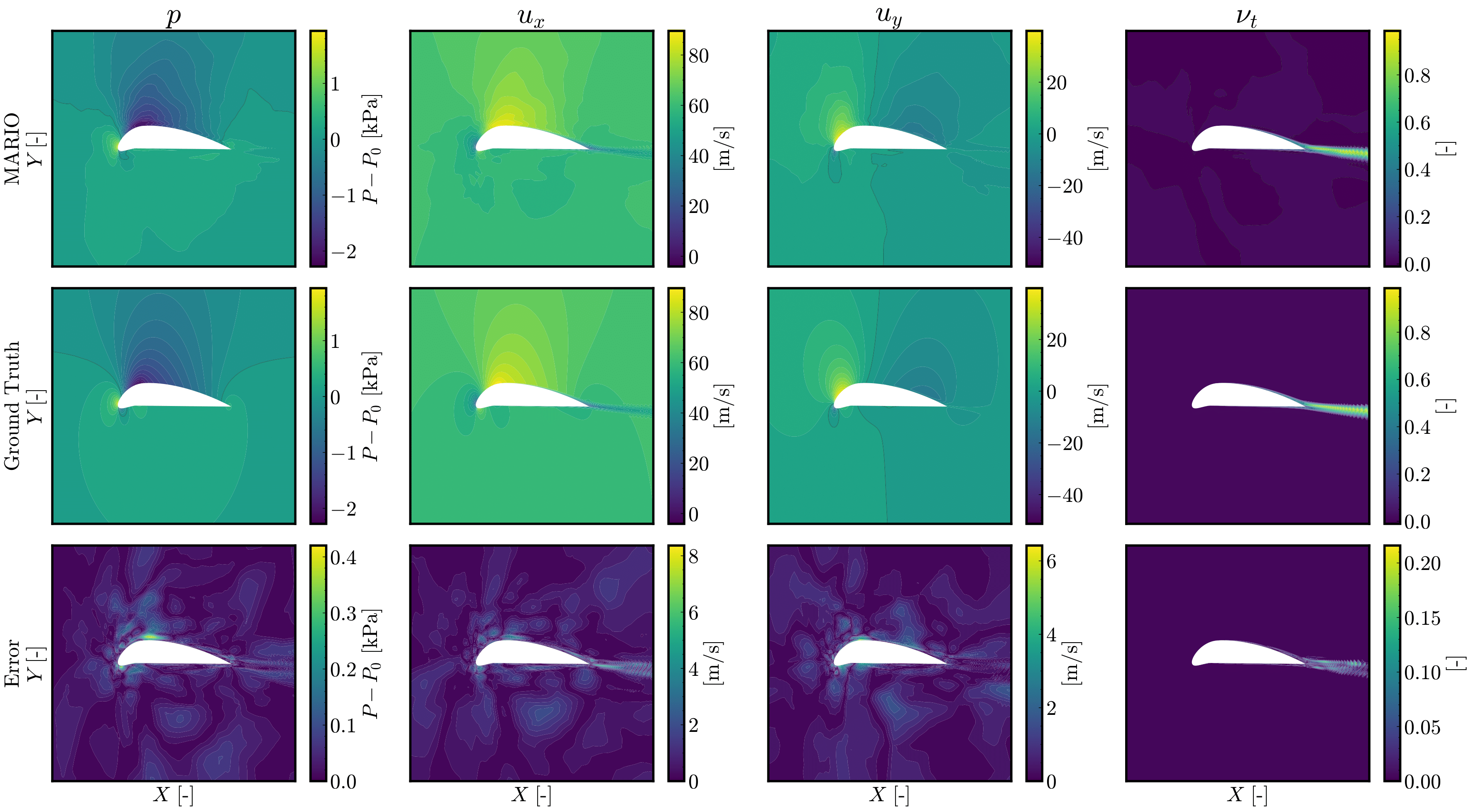}
    \caption{Comparison of all four predicted fields ($u_x$, $u_y$, $p$, $\nu_t$) for a challenging out-of-distribution geometry. First row: MARIO predictions; Second row: Ground truth CFD results; Third row: Absolute error contour plots. Operating conditions: $\alpha=0.196^\circ$, $M=0.18$.}
    \label{fig:fields_contour_96}
\end{figure}

\begin{table}[htbp]
\centering
\begin{tabular}{lcccc}
\toprule
\textbf{Model} & $\boldsymbol{C_D}$ & $\boldsymbol{C_L}$ & $\boldsymbol{\rho_D}$ & $\boldsymbol{\rho_L}$ \\
\midrule
MARIO & \textbf{0.794} & \textbf{0.115} & \textbf{0.102} & \textbf{0.997} \\
MLP* & \underline{2.950}  & 0.662  & -0.242  & 0.923  \\
PointNet* & 8.350  & 0.587  & -0.050  & 0.949  \\
GraphSAGE* & 3.504 & 0.385  & -0.139  & \underline{0.981}  \\
GUNet* & 6.871  & 0.418  & \underline{-0.095} & 0.976  \\
Transolver & 11.54 & \underline{0.206} & -0.292 & 0.956 \\
\bottomrule
\end{tabular}
\caption{\textbf{Comparison of drag coefficient ($C_D$), lift coefficient ($C_L$) Mean Relative Error, and their respective Spearman's rank correlations ($\rho_D$, $\rho_L$) for different models on the \textbf{scarce} task}. \textbf{Bold} values indicate the best performance (lowest for error metrics, highest for correlation metrics), while \underline{underlined} values indicate the second-best performance. For the models with *, values are taken from \cite{Bonnet_2023} under identical experimental settings.}
\label{tab:aero_coefficients}
\end{table}
First, the explicit geometry encoding mechanism allows complex airfoil shapes to be represented in a compact latent space, enabling efficient learning of geometric downstream effects on the physical fields. In order to demonstrate this point, we provide additional results in Appendix \ref{app:geometry_encoding} where the structure of the geometry latent space, resulting from the SDF encoding, is analyzed with respect to the main airfoil geometric features.
While Graph Neural Networks also have access to the SDF field, they lack an explicit mechanism to encode global geometric information, instead relying on local message passing, which struggles to capture the global influence of shape variations on flow patterns. It is worth noting that while Graph Neural Networks could potentially be augmented with the SDF encoding as an additional input feature, we did not explore this combination in the present study. 
Secondly, the neural field formulation benefits from resolution independence, allowing training on subsampled points while performing inference at any desired resolution without quality degradation. In contrast, graph-based approaches trained at lower resolutions must rely on domain subsampling during inference, leading to potential information loss and accumulation of errors, particularly in regions with complex flow structures.
Third, the combination of the boundary layer mask with Fourier feature encoding provides the model with the capability to accurately capture sharp gradients near the airfoil surface, which represents a critical region for aerodynamic performance prediction.
Table \ref{tab:training_inference_times} presents a comparison of training and inference times across selected models evaluated on the AirfRANS dataset. MARIO demonstrates competitive efficiency with training time comparable to other deep learning approaches. For inference, MARIO requires an initial geometry encoding step due to the optimization problem described in Section \ref{sec:Geometry Encoding}, followed by the physical field prediction. An important advantage of both MARIO and Transolver is their ability to directly process the full-resolution mesh during inference, as they are not dependent on the training resolution. In contrast, graph-based models must be iteratively applied to multiple subsamples (of similar resolution as the training) until the full mesh is covered, as described in the original paper \cite{Bonnet_2023}. Despite these differences in methodology, all surrogate models offer a massive speedup compared to traditional CFD simulations. Training times are generally similar across methods and depend on hyperparameters and network size.\\
\begin{table}[t]
\centering
\caption{\textbf{Comparison of training and inference times for different models on the AirfRANS scarce dataset}. Inference times are reported per sample on the full resolution mesh (CFD in seconds while surrogate in milliseconds). *Uses multiple inferences on batches of 32000 points as in the original paper. **MARIO requires two steps for inference on new geometries: geometry encoding (80ms) and physical field prediction (8ms).}
\label{tab:training_inference_times}
\begin{tabular}{lcc}
\toprule
\textbf{Model} & \textbf{Training Time} & \textbf{Inference Time } \\
\midrule
CFD & - &  1800 s \\
MARIO & 2h 30m & 88 ms** \\
GraphSAGE & 3h 25m & 130 ms* \\
GUNet & 3h 10m & 150 ms* \\
Transolver & 4h 50m & 8 ms \\
\bottomrule
\end{tabular}
\end{table}
The qualitative performance of the proposed model compared to baseline approaches is illustrated in Figures~\ref{fig:pressure_contour_82}, \ref{fig:fields_contour_99}, and \ref{fig:fields_contour_96}. Figure~\ref{fig:pressure_contour_82} provides a direct comparison between the pressure field predictions of MARIO and Transolver against the ground truth. The surface pressure coefficient plot clearly demonstrates the superior accuracy of the neural field approach, particularly in capturing the pressure peak on the airfoil's upper surface.
Figure~\ref{fig:fields_contour_99} presents predictions for all four output fields on a test sample with an in-distribution airfoil shape. As expected, most prediction errors are concentrated in regions of high gradients near the airfoil surface and in the wake region for turbulent viscosity. These areas pose modeling challenges even for conventional CFD approaches, as they require extremely fine mesh resolution to be accurately resolved. Despite these inherent difficulties, the model maintains good prediction quality throughout the domain.
The robustness of the approach is further tested in Figure~\ref{fig:fields_contour_96}, which shows predictions for an out-of-distribution airfoil geometry featuring an unusual leading edge curvature. This represents a particularly challenging test case, as the model must generalize to geometric features not well-represented in the training data. While prediction quality naturally decreases, with some noise becoming apparent in the predicted fields, the major flow structures remain well-captured. The most significant discrepancy appears in the pressure prediction at the suction peak, where the extreme local curvature induces very steep pressure gradients that are difficult to generalize from the training distribution. Nevertheless, the overall flow pattern prediction remains reliable.

\subsubsection{Effect of Scale Encoding and Boundary Layer Mask}

\begin{figure}[h]
    \centering
    \includegraphics[width=\textwidth]{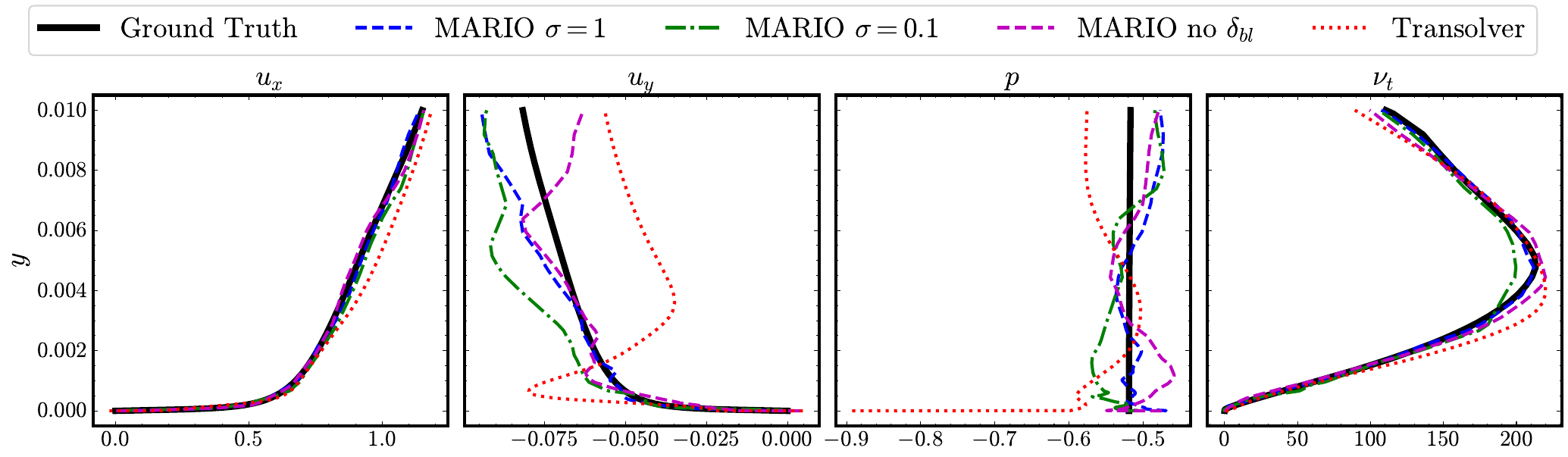}
    \caption{Comparison of velocity, pressure, and turbulent viscosity profiles in the boundary layer region at x=0.5c for different MARIO model variants. $\alpha=7.91 ^\circ$, $M=0.16$}
    \label{fig:bl_05}
\end{figure}

The impact of different model configurations on aerodynamic coefficient predictions is presented in Table~\ref{tab:aero_coefficients_mario_variants}. Three variants are compared: the standard configuration with Fourier feature encoding at $\sigma=1$, a variant with reduced encoding frequency ($\sigma=0.1$), and a variant without the boundary layer mask ($\sigma_{bl}$). 
The results point out the crucial role of both the frequency scale in the Fourier feature encoding and the boundary layer mask. While both $\sigma$ values provide similar performance on lift coefficient prediction, the higher frequency encoding ($\sigma=1$) yields significantly better drag coefficient predictions. This can be attributed to the drag coefficient's greater sensitivity to accurate resolution of pressure gradients and flow separation, which require capturing higher-frequency components of the solution.
The most dramatic effect is observed when removing the boundary layer mask, where drag coefficient errors increase nearly sixfold. This underscores the importance of directing the model's attention to the near-wall region where the steepest gradients occur. Interestingly, while drag prediction deteriorates substantially without the boundary layer mask, lift prediction remains relatively robust, as lift forces are mainly caused by potential flow effects. This is especially true at higher Reynolds numbers, lower angles of attack and in the absence of flow separation, as the de-cambering effect due to boundary layer is less prominent.
Figure~\ref{fig:bl_05} provides further insight into how these model variants perform within the boundary layer region. The standard model with $\sigma=1$ most accurately captures the velocity profile across the boundary layer, particularly the steep gradient near the wall. The higher-frequency encoding allows the model to represent these sharp transitions without introducing spurious oscillations. In contrast, the variant without the boundary layer mask exhibits noticeable oscillations in the predicted velocity field. Without incorporating the boundary layer input, the network is forced to explain the entire range of gradients in the flow field using a single set of global weights. Consequently, in order to capture the steep gradients near the wall, the network adopts higher sensitivity which may result in spurious oscillations in regions where the solution is inherently smooth.
By integrating the BL function \(\sigma_{bl}(x)\) as an additional input, the network is conditioned to differentiate between regions requiring distinct sensitivities. Specifically, in regions close to the wall where \(\sigma_{bl}(x)\) is high, the network is allowed to exhibit a high local Lipschitz constant to accurately capture the sharp gradients. Conversely, in areas farther from the wall where \(\sigma_{bl}(x)\) is low or zero, the network naturally maintains a lower local Lipschitz constant, yielding a smoother response. In effect, the BL function \(\sigma_{bl}(x)\) enables the network to effectively \emph{modulate} its local sensitivity as a function of position in the domain, capturing the essential features of the flow field without overfitting in smooth regions.

\begin{table}[t]
\centering
\begin{tabular}{lcccc}
\toprule
\textbf{Model} & $\boldsymbol{C_D}$ & $\boldsymbol{C_L}$ & $\boldsymbol{\rho_D}$ & $\boldsymbol{\rho_L}$ \\
\midrule
MARIO $\sigma=1$ & \textbf{0.794} & \textbf{0.115} & \textbf{0.102} & \textbf{0.997} \\
MARIO $\sigma=0.1$ & 0.946 & 0.124 & 0.167 & 0.997 \\
MARIO (no $\delta_{bl}$) & 4.780 & 0.151 & -0.074 & 0.995 \\
\bottomrule
\end{tabular}
\caption{\textbf{Effect of model configuration on aerodynamic coefficient predictions.} $C_D$ and $C_L$ values represent Mean Relative Error (\%) for drag and lift coefficients, while $\rho_D$ and $\rho_L$ are Spearman's rank correlation coefficients. Bold values indicate best performance.}
\label{tab:aero_coefficients_mario_variants}
\end{table}

\subsection{NASA Common Research Model}
\label{subsec:result_crm}

\begin{figure}[h]
    \centering
    \begin{subfigure}[b]{0.48\textwidth}
        \centering
        \includegraphics[width=\textwidth]{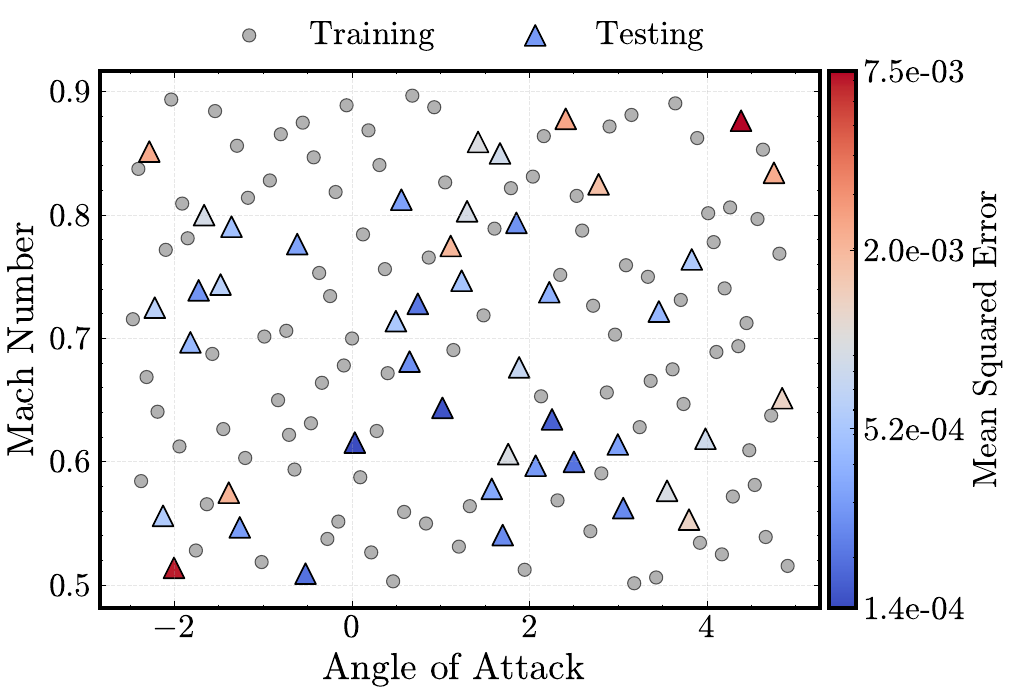}
    \end{subfigure}
    \hfill
    \begin{subfigure}[b]{0.48\textwidth}
        \centering
        \includegraphics[width=\textwidth]{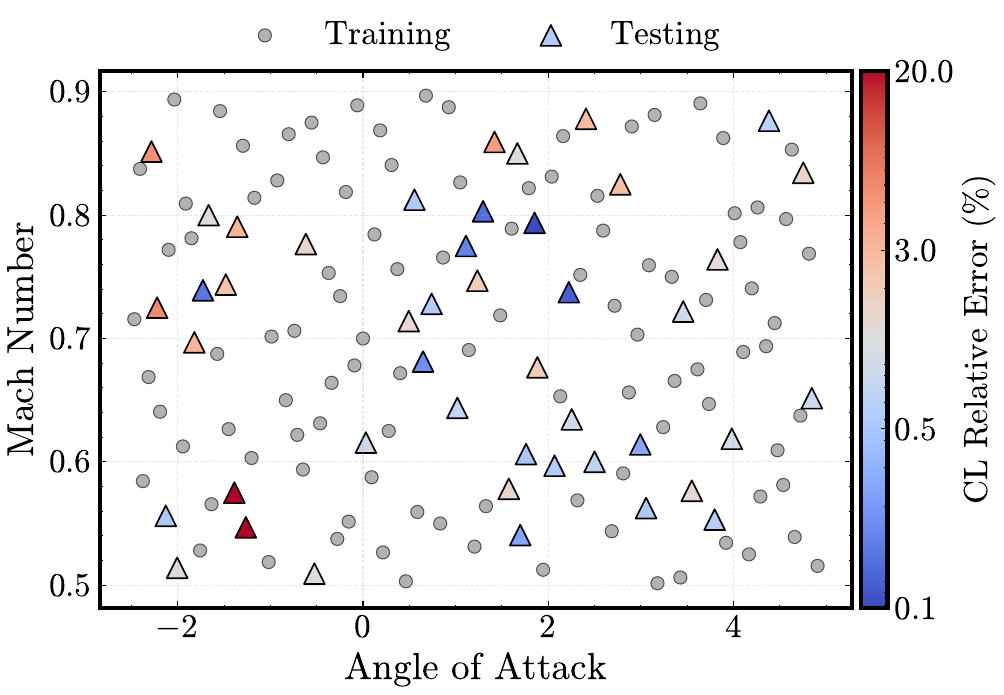}
    \end{subfigure}
    \caption{Scatter plots of prediction errors across the Mach number and angle of attack parameter space. Left: Mean Squared Error (MSE) on pressure coefficient ($C_p$). Right: Mean Relative Error on lift coefficient ($C_L$). Color-coded triangles represent test samples. Grayscale triangles indicate the training dataset distribution. Note that while Mach number and angle of attack are displayed, other parameters such as control surface deflections are also varied in the dataset.}
    \label{fig:scatter_plots}
\end{figure}

The performance of the MARIO framework on the NASA Common Research Model dataset is summarized in Table \ref{tab:crm_performance}. The model achieves good prediction accuracy on the pressure coefficient of the test set samples. The mean relative error for the lift coefficient predictions is 3.12\% on average. The highest errors are observed for points in the negative low Mach number and angle of attack ranges, as shown in Figure \ref{fig:scatter_plots}. Notably, for cases within the training parameters ranges, the prediction accuracy improves significantly, with lift coefficient errors below 0.5\%.
Error analysis reveals distinct patterns across the parameter space. As illustrated in Figure \ref{fig:scatter_plots}, prediction errors are primarily concentrated at the extremes of the parameter space, specifically at combinations of the lowest Mach number with the lowest angle of attack, and the highest Mach number with the highest angle of attack. This relates to the difficulty of the surrogate model when extrapolating beyond the training distribution, where the flow physics can change significantly with small parameter variations. Figure \ref{fig:scatter_plots} shows that lift coefficient errors similarly increase in the low Mach number and negative angle of attack range, where the integrated effects of pressure distribution inaccuracies become more pronounced.

\begin{table}[t]
    \centering
    \caption{\textbf{MARIO Performance on NASA CRM Dataset.} MSE: Mean Squared Error on pressure coefficient ($C_p$, $\times 10^{-3}$); MRE: Mean Relative Error on lift coefficient ($C_L$, \%); Training time for 10,000 epochs on single GPU; Inference time for full surface mesh (454,404 nodes).}
    \label{tab:crm_performance}
    \begin{tabular}{cccc}
    \toprule
    $\boldsymbol{C_p}$ \textbf{(MSE)} & $\boldsymbol{C_L}$ \textbf{MRE} & \textbf{Training (h)} & \textbf{Inference (s)} \\
    \midrule
    0.76 & 3.12\% & 3.98 & 0.068 \\
    \bottomrule
    \end{tabular}
\end{table}

\begin{figure}[h]
    \centering
    \includegraphics[width=\textwidth]{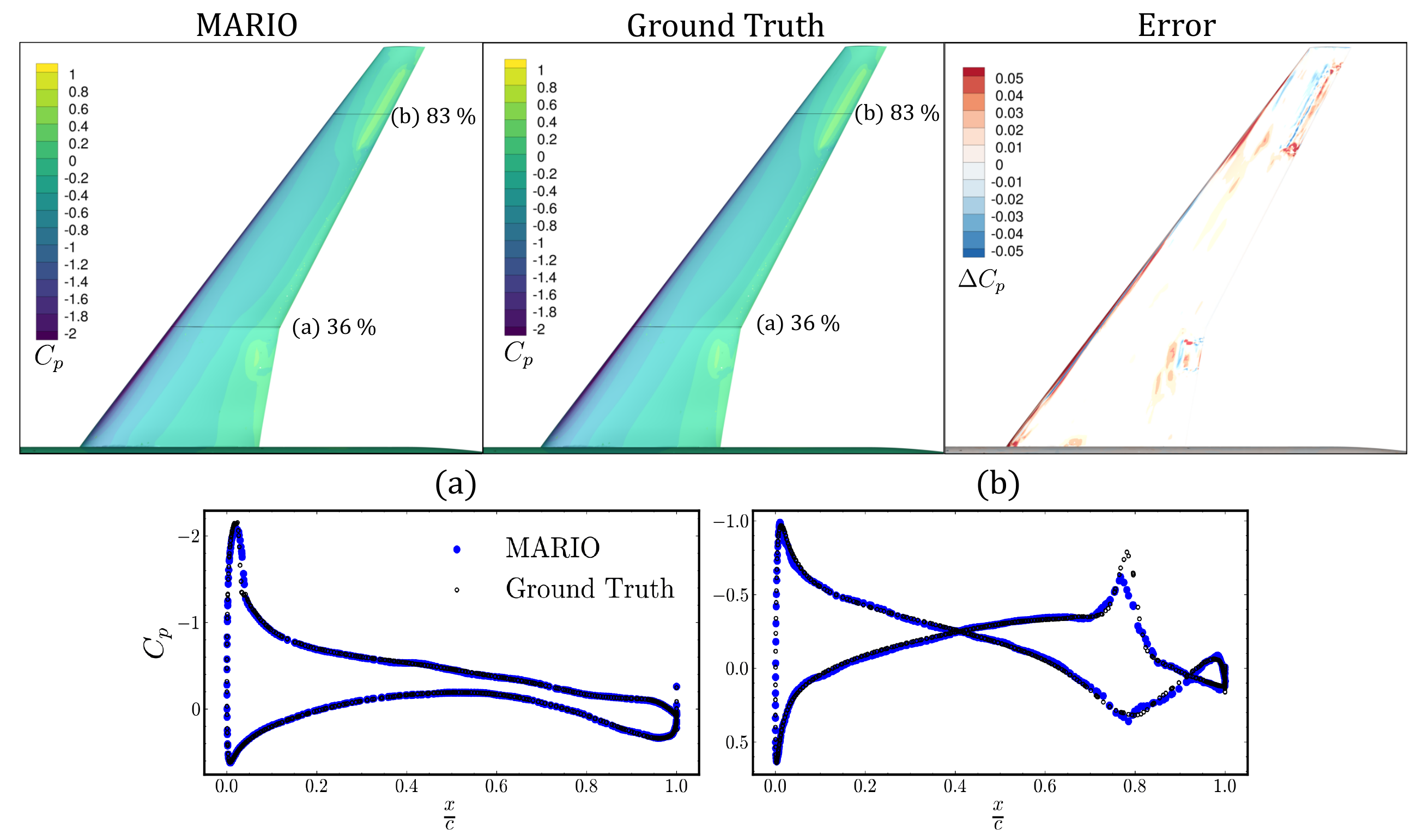}
    \caption{Top: pressure coefficient ($C_p$) fields predictions and targets with spatial errors on the wing for NASA CRM test case. Bottom: $C_p$ vs choordwise position at two distinct spanwise locations. Operating conditions: $\alpha = 3.98^\circ$, $M = 0.62$, $\phi_{outAil} = -19.52^\circ$, $\phi_{inAil} = -14.16^\circ$, $\phi_{el} = 1.57^\circ$, $\phi_{htp} = 1.46^\circ$. }
    \label{fig:nasa_contour_36}
\end{figure}

\begin{figure}[h]
    \centering
    \includegraphics[width=\textwidth]{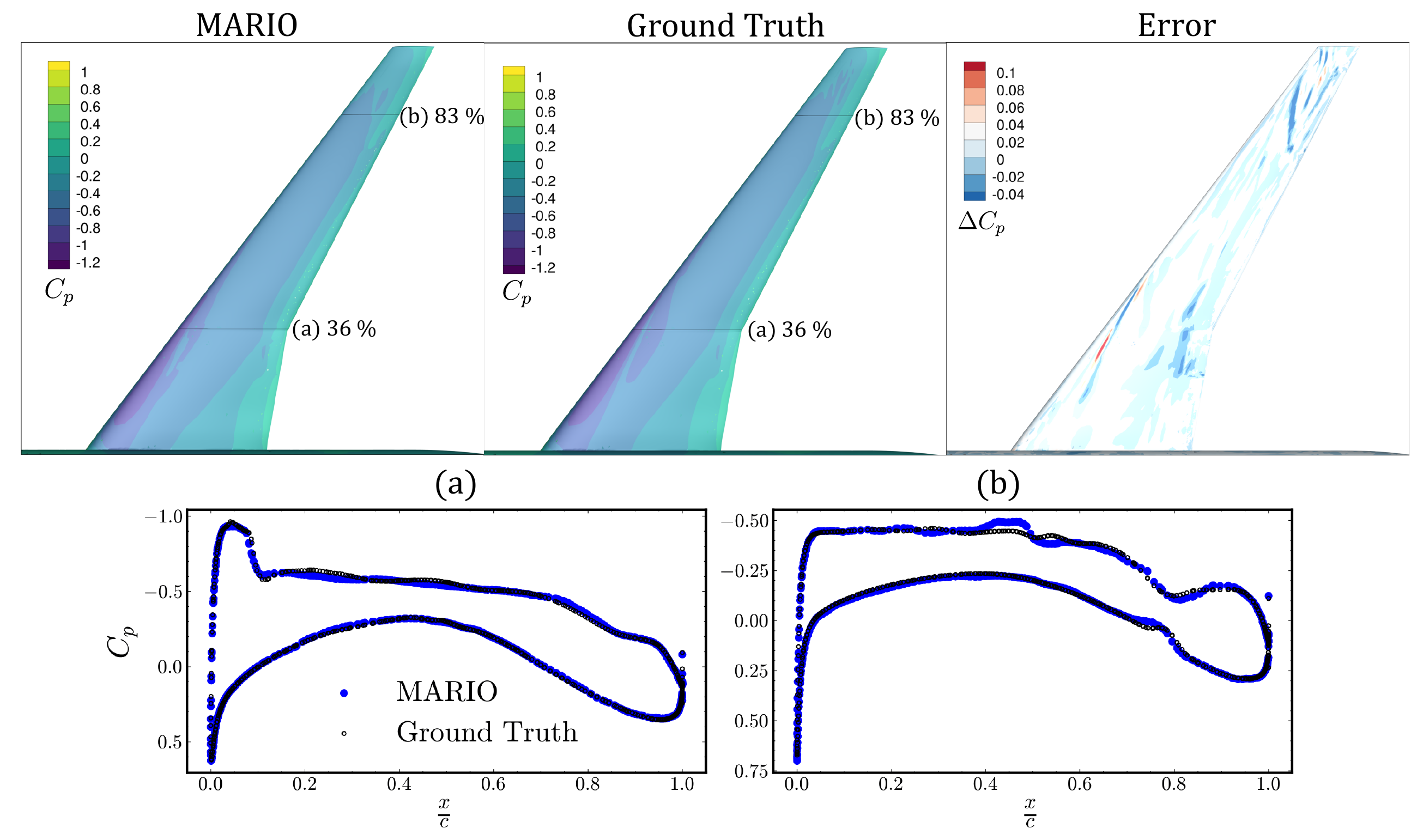}
    \caption{Top: pressure coefficient ($C_p$) fields predictions and targets with spatial errors on the wing for NASA CRM test case. Bottom: $C_p$ vs choordwise position at two distinct spanwise locations. Operating conditions: $\alpha = 1.30^\circ$, $M =0.80 $, $\phi_{outAil} = -3.92^\circ$, $\phi_{inAil} = 4.60^\circ$, $\phi_{el} =-7.19 ^\circ$, $\phi_{htp} = -1.27^\circ$. }
    \label{fig:nasa_contour_15}
\end{figure}

\begin{figure}[h]
    \centering
    \includegraphics[width=\textwidth]{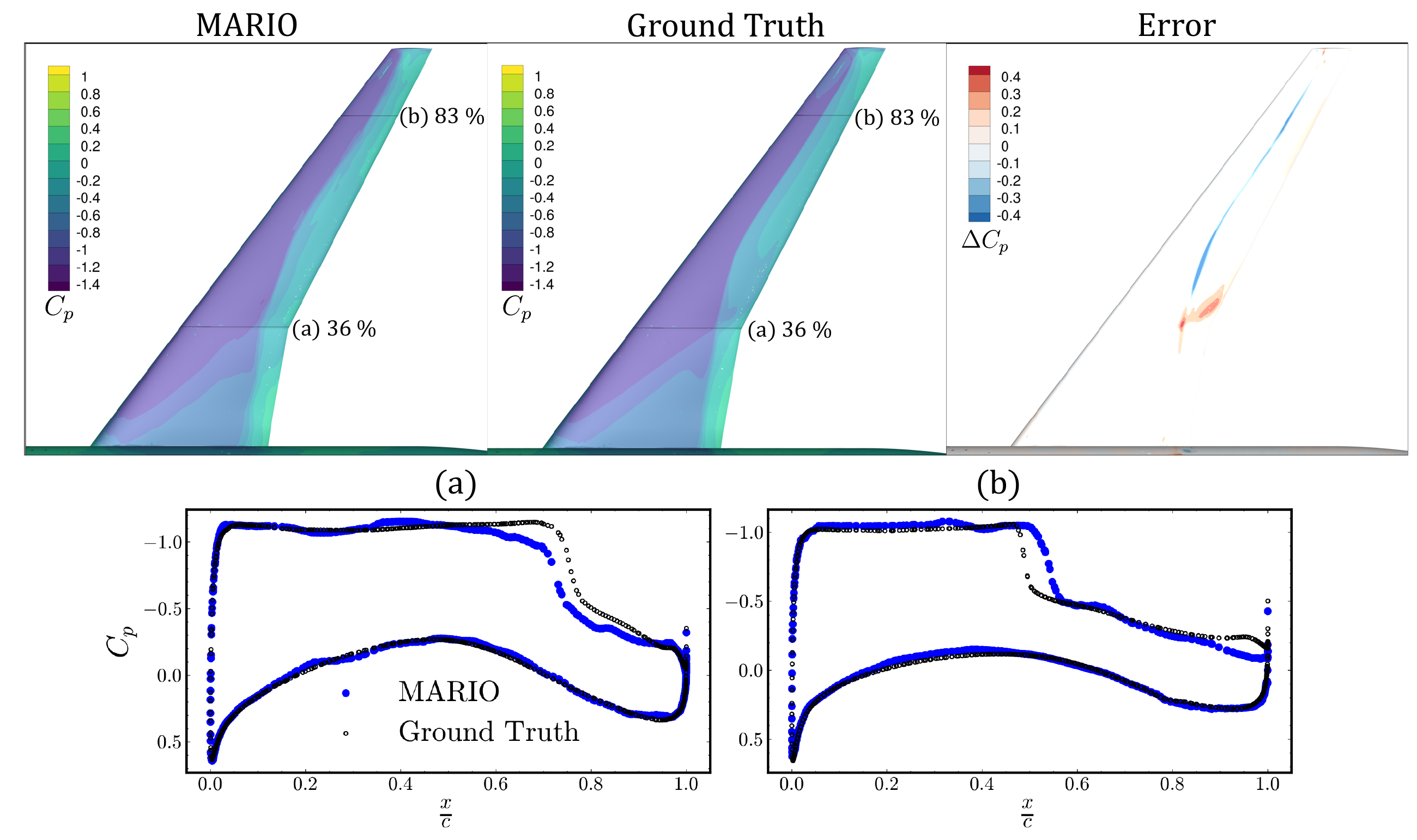}
    \caption{Top: pressure coefficient ($C_p$) fields predictions and targets with spatial errors on the wing for NASA CRM test case. Bottom: $C_p$ vs choordwise position at two distinct spanwise locations. Operation conditions: $\alpha = 4.38^\circ$, $M =0.88 $, $\phi_{outAil} = 1.65^\circ$, $\phi_{inAil} = 2.27^\circ$, $\phi_{el} =-9.65 ^\circ$, $\phi_{htp} = -1.74^\circ$. }
    \label{fig:nasa_contour_extreme}
\end{figure}

Figures \ref{fig:nasa_contour_36}, \ref{fig:nasa_contour_15}, \ref{fig:nasa_contour_extreme} provide qualitative insights into the model's predictive performance of the surface pressure distribution across three different operating conditions and physical regimes: entirely subsonic flow over the upper surface,  weak shock wave formation, strong shock wave over the entire wing span respectively. Figure \ref{fig:nasa_contour_36} shows a case at $M=0.62$ with significant control surface deflections ($\phi_{outAil} = -19.52^\circ$, $\phi_{inAil} = -14.16^\circ$). The model accurately captures the main pressure patterns, particularly the interaction between outboard aileron deflection and leading edge vortices.  Prediction errors are primarily concentrated at the leading edge suction peak, which can be attributed to the higher absolute pressure values in this region rather than a failure to capture the underlying physics.
Figure \ref{fig:nasa_contour_15} demonstrates the model's capability in higher Mach number conditions ($M=0.80$). At this flight condition, local supersonic regions form on the upper surface of the wing, resulting in characteristic pressure jumps. Despite these more complex flow features, the predictions remain accurate in both magnitude and spatial distribution. The model correctly identifies the location of pressure jumps and captures the pressure field variations around deflected control surfaces.

Figure \ref{fig:nasa_contour_extreme} presents a challenging case at $M=0.87$ and $\alpha=4.2^\circ$, representing an extrapolation scenario at the edge of the training distribution. While the overall pressure field distribution is still reasonably captured across most of the aircraft surface, notable discrepancies appear in regions of strong pressure gradients. Particularly, the shock location is predicted further upstream, closer to the wing root, and further downstream, closer to the wingtip, compared to the reference CFD solution. This highlights a fundamental challenge in both surrogate modeling and traditional CFD of compressible flows, where small parameter variations can lead to significant changes in shock position and strength. Indeed, even high-fidelity CFD simulations frequently exhibit discrepancies with experimental data regarding precise shock locations, as these flow features are particularly sensitive to small perturbations in boundary conditions, numerical methods, and turbulence modeling approaches. 
The model demonstrates good computational efficiency. Training for 10,000 epochs takes approximately 4 hours on a single GPU. This is partly achieved through the training strategy of subsampling a much smaller subset of points, leveraging the neural field's resolution independence. Inference time is less than one-tenth of a second for the full mesh containing 454,404 nodes, providing a massive speedup (6 orders of magnitude) compared to high-fidelity CFD simulations that require approximately 19 hours on 64 CPUs as reported in the original paper \cite{doi:10.2514/1.J061234}.
It is worth noting that this comparison does not account for the initial training time and model selection process, so the computational advantages are most significant when the model is already trained and multiple evaluations are needed, such as in design optimization scenarios. Additionally, the comparison is not iso-accuracy, as there remains a gap between the surrogate predictions and the high-fidelity CFD results that served as training data.

\section{Conclusions}
\label{sec:conclusions}

In this work, we introduced MARIO (Modulated Aerodynamic Resolution Invariant Operator), a surrogate modeling framework based on Neural Fields for aerodynamic simulations. The approach utilizes the discretization-invariant nature of Neural Fields to handle both parametric and non-parametric geometric variations while reducing computational costs compared to traditional CFD solvers. For non-parametric geometric variations, MARIO employs an encoder-decoder architecture that first encodes the geometry through its Signed Distance Function before predicting physical fields. In contrast, for parametric variations, a simpler decoder-only architecture directly incorporates the explicit shape parameters. This architectural flexibility allows MARIO to scale effectively from 2D airfoil flows to complex 3D aircraft configurations without substantial modifications.
MARIO offers strong computational advantages in both training and inference phases. The discretization-invariant formulation enables training on significantly downsampled meshes (using only 2-10\% of the original mesh points), dramatically reducing memory requirements compared to graph-based approaches that typically process the full mesh structure. 
The presented experimental evaluation spans two industrially relevant test cases: the 2D AirfRANS airfoil benchmark and the 3D NASA Common Research Model, providing a comprehensive assessment of the proposed approach. Both datasets feature large-scale meshes but differ in dimensionality, flow regime, and geometric variation type. On the AirfRANS dataset, an extensive benchmark comparison against state-of-the-art approaches based on Graph Neural Networks and operator learning is provided. The results demonstrate that MARIO achieves approximately an order of magnitude improvement in prediction accuracy across all output fields compared to the strongest baselines, in the data-scarce regime.

Futhermore, the analysis of model components for the AirfRANS dataset shows the importance of both Fourier feature encoding and the boundary layer mask in capturing steep flow gradients near the airfoil surface. While higher frequency encoding ($\sigma=1$) provides better drag coefficient predictions, the boundary layer mask proved essential for directing the model's attention to the near-wall region, with its removal resulting in a sixfold increase in drag prediction errors. 
For the NASA CRM dataset, the model demonstrates good accuracy in predicting surface pressure distributions on a full aircraft configuration with control surface deflections, achieving a mean relative error of 3.12\% on lift coefficient predictions. For in-distribution cases, the error on lift coefficient drops below 0.5 \%, showing excellent predictive capability when interpolating within the training domain. The highest errors were observed at the boundaries of the parameter space, particularly when extrapolating to combinations of parameters not well-represented in the training data.
Importantly, these results are achieved in realistic data-scarce scenarios, using only 200 training samples for AirfRANS and 150 for the NASA CRM, reflecting the practical constraints often encountered in industrial settings where high-fidelity simulations are computationally expensive.

\paragraph{Limitations and future work:} Despite the promising results, several limitations remain. The approach still shows reduced accuracy when extrapolating to configurations outside the training distribution, particularly at the corners of the parameter space. This suggests the need for active learning strategies to optimize sampling in regions of high prediction uncertainty. Additionally, while we demonstrated performance on both 2D and 3D configurations, further work is needed to address multi-component aircraft geometries with parts at different length scales and complex interactions between components. Future research should also explore extensions to unsteady flows, where the temporal dynamics presents additional modeling challenges not addressed in the current framework.
As the field of machine learning for fluid dynamics continues to evolve, standardized benchmarking across diverse datasets and scenarios will be essential for objectively evaluating and comparing surrogate modeling approaches. Toward this end, MARIO has already demonstrated strong performance beyond the experiments presented in this paper, securing third place at the NeurIPS ML4CFD challenge \cite{yagoubi2024neurips}, which featured the AirfRANS dataset under extreme data scarcity conditions (100 training samples). Furthermore, MARIO has been evaluated on the recently proposed Plaid benchmark \cite{casenave2025physicslearningaidatamodelplaid} hosted on Hugging Face, showing competitive performance across a variety of test cases.

\section*{Acknowledgements}
This work was was supported by Agence Nationale de la Recherche (ANRT), through  CIFRE PhD Fellowship sponsored by Airbus Operations SAS and ISAE-Supaero. 
We also thank Philipp Bekemeyer and Derrick Hines from DLR (German Aerospace Center) for kindly sharing the CFD data on the NASA CRM test case.







\bibliographystyle{elsarticle-num-names}
\bibliography{sample.bib}







\appendix

\section{Analysis of AirfRANS Geometry Representations}
\label{app:geometry_encoding}
\begin{figure}[h]
    \centering
    \includegraphics[width=\linewidth]{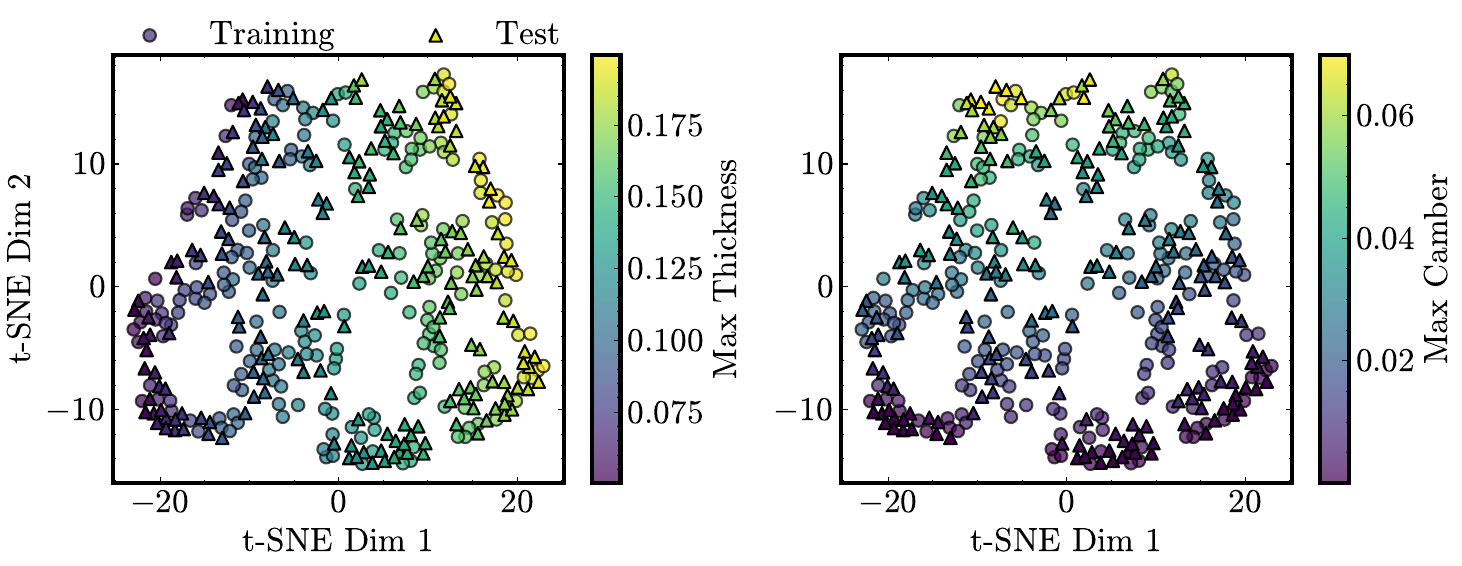}
    \caption{t-SNE visualization of the 8-dimensional latent geometry codes learned from the AirfRANS dataset. Color indicates maximum thickness value (left) and maximum camber value (right).}
    \label{fig:tsne_geometry_codes}
\end{figure}

In this appendix, we analyze the geometric representations learned by the MARIO framework for the AirfRANS dataset. As described in Section~\ref{sec:Geometry Encoding}, the auto-encoding process extracts 8-dimensional latent codes from the Signed Distance Function (SDF) fields that characterize each airfoil geometry. These latent codes subsequently serve as conditioning variables for the flow field predictions. To examine the structure of these learned representations, t-SNE dimensionality reduction is applied to the 8-dimensional latent codes, yielding the two-dimensional embedding shown in Figure~\ref{fig:tsne_geometry_codes}. Each point in this visualization corresponds to an individual airfoil geometry, color-coded according to its maximum camber value. Analysis reveals that the embedding exhibits meaningful organization with respect to key aerodynamic parameters: horizontal organization approximately corresponds to iso-camber contours, while vertical correlates with iso-thickness levels. This structure demonstrates that the neural encoding effectively captures the principal geometric features relevant to aerodynamic performance, despite not being explicitly trained to preserve these specific parameters. The latent representations are compact and well-distributed around the origin, without forming discrete clusters.
To quantitatively assess the effectiveness of these learned representations, we compare the predictive performance of MARIO when using the learned latent codes versus a direct thickness and camber distribution approach. In the latter implementation, the geometry is parameterized by explicitly measuring thickness and camber distributions at 8 fixed chordwise locations. While this approach provides an intuitive, physically-based description in the 2D airfoil context, it lacks the flexibility required for more complex configurations such as 3D geometries or multi-element airfoils, and cannot capture fine geometric details that may influence the flow field.

Table~\ref{tab:downstream_comparison} presents the comparative evaluation results for both representations. 

\begin{table}[h]
\centering
\caption{\textbf{Comparison of different geometry representations on field prediction performance}. Values shown are absolute MSE errors on the test set (standardized outputs). Error values are scaled as follows: $\bar{u}_x$, $\bar{u}_y$, and $\bar{p}$ are multiplied by $10^{-2}$; $\nu_t$ and $\bar{p}_s$ are multiplied by $10^{-1}$.}
\label{tab:downstream_comparison}
\begin{tabular}{lccccc}
\toprule
\textbf{Representation} & $\boldsymbol{\bar{u}_x}$ & $\boldsymbol{\bar{u}_y}$ & $\boldsymbol{\bar{p}}$ & $\boldsymbol{\nu_t}$ & $\boldsymbol{\bar{p}_s}$ \\
\midrule
SDF Encoding (8D) & 0.152 & 0.113 & 0.24 & 0.096 & 0.27 \\
Thickness-Camber  (16 D) & 0.121 & 0.094 & 0.22 & 0.108 & 0.25 \\
\bottomrule
\end{tabular}
\end{table}

The comparative evaluation reveals similar performance between the 8-dimensional SDF-based latent encoding and the 16-dimensional thickness-camber representation. While the thickness-camber approach shows slightly lower errors across velocity fields predictions, the differences are marginal—typically less than 10-15\% relative improvement. This suggests that the learned SDF encoding captures geometric information as effectively as an explicitly engineered geometric parameterization and with fewer parameters.

\end{document}